\documentclass[aps,prd,groupedaddress,twocolumn,nofootinbib]{revtex4}
\usepackage{amsmath,graphicx}
\usepackage{hyperref}
\usepackage{cleveref}

\usepackage{natbib}
\usepackage{mathrsfs}

\begin{document}

\title{Generic rotating regular black holes in general relativity coupled\\ to nonlinear electrodynamics}

\author{Bobir Toshmatov$^{1,2}$}
\email{bobir.toshmatov@fpf.slu.cz}

\author{Zden\v{e}k Stuchl\'{i}k$^{1}$}
\email{zdenek.stuchlik@fpf.slu.cz}

\author{Bobomurat Ahmedov$^{3,2}$}
\email{ahmedov@astrin.uz}

\affiliation{%
$^{1}$ Institute of Physics and Research Centre of Theoretical Physics and Astrophysics, Faculty of Philosophy \& Science, Silesian University in Opava, Bezru\v{c}ovo n\'{a}m\v{e}st\'{i} 13,  CZ-74601 Opava, Czech Republic\\
$^{2}$ Institute of Nuclear Physics, Ulughbek, Tashkent 100214, Uzbekistan\\
$^{3}$ Ulugh Beg Astronomical Institute, Astronomicheskaya 33, Tashkent 100052, Uzbekistan}

\begin{abstract}
We construct regular rotating black hole and no-horizon spacetimes based on the recently introduced spherically symmetric generic regular black hole spacetimes related to electric or magnetic charge under nonlinear electrodynamics coupled to general relativity that for special values of the spacetime parameters reduce to the Bardeen and Hayward spacetimes. We show that the weak and strong energy conditions are violated inside the Cauchy horizons of these generic rotating black holes. We give the boundary between the rotating black hole and no-horizon spacetimes and determine the black hole horizons and the boundary of the ergosphere. We introduce the separated Carter equations for the geodesic motion in these rotating spacetimes. For the most interesting new class of the regular spacetimes, corresponding for magnetic charges to the Maxwell field in the weak field limit of the nonlinear electrodynamics, we determine the structure of the circular geodesics and discuss their properties. We study the epicyclic motion of a neutral particle moving along the stable circular orbits around the "Maxwellian" rotating regular black holes. We show that epicyclic frequencies measured by the distant observers and related to the oscillatory motion of the neutral test particle along the stable circular orbits around the rotating singular and regular Maxwellian black holes are always smaller than ones in the Kerr spacetime.
\end{abstract}

%\pacs{04.50.Kd, 04.70.-s, 04.25.-g}

\maketitle

\section{Introduction}\label{sec-intr}

In the standard general relativity black hole solutions contain a physical singularity with diverging Riemann tensor components, considered as a region governed by quantum gravity overcoming this internal defect of general relativity. On the other hand, families of regular black hole solutions of Einstein's gravity have been found where the physical singularity is eliminated. Of course, such solutions cannot be vacuum solutions of the Einstein equations, but contain necessarily a properly chosen additional field guaranteeing violation of the energy conditions related to the existence of physical singularities \cite{Haw-Elli:1973:LargeScaleStructure:}. Another way of obtaining a singularity free black hole solution is modification of the gravitational law, as in the Kehagias-Sfetsos black hole solutions \cite{Keh-Sfe:2009:PhysLetB:} in the modified Ho\v{r}ava quantum gravity \cite{Hor:2009:PHYSR4:,Hor:2009:PHYSRL:}.

The regular black hole solution with a magnetic charge has been proposed by Bardeen \cite{Bardeen:1968}. It was shown that the magnetic charge has to be related to a nonlinear electrodynamics \cite{AyB-Gar:2000:PhysLetB:}. The other solution of the combined Einstein and nonlinear electrodynamic equations has been introduced by Ayon-Beato and Garcia \cite{AyB-Gar:2000:PhysLetB:,AyB-Gar:1998:PhysRevLet:,AyB-Gar:1999:PhysLetB:,AyB-Gar:1999:GenRelGrav:}. A different approach to the regular black hole solutions has been applied by Hayward \cite{Hay:2006:PhysRevLet:}. Modification of the mass function in the Bardeen and Hayward solutions and inclusion of the cosmological constant can be found in the new solutions of Neves and Saa \cite{Nev-Saa:2014:arXiv:1402.2694:,Dym-Gal:2016:CLAQG:,Nev:PRD:2015}. The regular black hole solution in the $f(T)$ gravity with nonlinear electrodynamics has been found in \cite{Jun-Rod-Hou:2015:JCAP:}. All the subtleties of the nonlinear effects were discussed in \cite{Bronnikov:2006,Azreg-AinouPLB:2014,Dyn:2015:CLAQG:,Dym:2016:IJMPD:}. Rotating regular black hole solutions have been introduced in \cite{Mod-Nic:2010:PHYSR4:,Azreg-Ainou2014PRD,Bam-Mod:2013:PhysLet:,Tos-Abd-Ahm-Stu:2014:PHYSR4:,Gho:2015:EPJC:,Ami-Gho:2015:JHEP:}. Similarly to the Kerr or Reissner-Nordstr\"{o}m solutions of the standard general relativity that describe both black holes and naked singularities, the regular solutions describe both the black hole and no-horizon spacetimes demonstrating strong gravity effects \cite{DeFeliceAA:1974,Stu:1980:BAC:,Stu-Hle:2002:ActaPhysSlov:,Pug-Que-Ruf:2011:PHYSR4:, Pug-Que-Ruf:2011:PHYSR4b:,Stu-Sche:2013:CLAQG:,Stu-Sche:2015:IJMPD:,Sche-Stu:2015:JCAP:}.

Properties of the geodesic motion in the field of regular black holes have been recently discussed in \cite{Gar-Hac-Kun-Lam:2015:,PatilPRD:2012,Tos-Abd-Ahm-Stu:2015:APSS4:,AbdujabbarovPRD2016}. A detailed discussion of the circular geodesics of the regular Bardeen and Ayon-Beato-Garcia (ABG) black hole and no-horizon spacetimes and its implication to simple optical phenomena can be found in \cite{Stu-Sche:2015:IJMPD:}. Moreover, scalar, electromagnetic and gravitational perturbations of the regular black holes in nonlinear electrodynamics, their quasinormal modes and stabilities have been discussed in several works, see for instance~\cite{FernandoPRD:2012,FlachiPRD:2013,LiPRD:2013,ToshmatovPRD:2015,MorenoPRD:2003,ChaverraPRD:2016}. Note that the geodesic structure of the regular black holes outside the horizon is similar to those of the Schwarzschild or Reissner-Nordstr\"{o}m (RN) black hole spacetimes, but under the inner horizon, no circular geodesics can exist. The geodesic structure of the no-horizon spacetimes is similar to those of the naked singularity spacetimes of the RN type, or the Kehagias-Sfetsos type \cite{AtamurotovAPSS2013,Stu-Sche:2014:CLAQG:,HakimovPRD:2013,Stu-Sche-Abd:2014:PHYSR4:,ToshmatovAPSS:2016} representing a widely discussed asymptotically flat solution of modified Ho\v{r}ava quantum gravity \cite{Keh-Sfe:2009:PhysLetB:,Hor:2009:PHYSR4:,Hor:2009:PHYSRL:}.

A fundamental difference between the naked singularity and regular spacetimes is related to the central region. In the regular Bardeen and ABG spacetimes the metric is de Sitter like at the central region \cite{Stu-Sche:2015:IJMPD:}, just in the spirit of the original ideas on the nonsingular black hole spacetimes presented in \cite{Sak:1966:JETP:,Gli:1966:JETP:}. In the no-horizon spacetimes this effect has an extraordinary influence on the character of null geodesics and related optical phenomena \cite{Sche-Stu:2015:JCAP:}.

Quite recently, a new class of generic regular spherically symmetric solutions of the standard general relativity coupled to nonlinear electrodynamics has been found for both magnetically and electrically charged gravitational objects \cite{Fan-Wan:2016:arxiv1610.02636:}. For special choice of the spacetime parameters, this generic solution reduces to the Bardeen and Hayward spacetimes. There is also a new interesting special class of this solution corresponding to the Maxwell field in the weak field limit of the nonlinear electrodynamics -- this is demanding special attention.

Here using the Newman-Janis algorithm we construct rotating regular black hole and no-horizon spacetimes related to the new generic class of these
spacetimes. We discuss their basic properties and give the Carter equations of the geodesic motion. Then we give the circular geodesics of the special new class with the magnetic Maxwell limit, discuss their properties, and compare them to the well-known Bardeen spacetime. We also give the frequencies of the radial and vertical epicyclic motion, as these frequencies could be relevant in explaining the high-frequency quasiperiodic oscillations \cite{Tor-Abr-Klu-Stu:2005:ASTRA:,Stu-Kot:2009:GenRelGrav:,Stu-Sche:2012:CLAQG:} and could thus serve as a strong test of relevance of the regular solutions.

The paper is organized as follows: in Sec.~\ref{sec-nj} we briefly describe the Newman-Janis algorithm. In Sec.~\ref{sec-rbh} the generic rotating black hole solution is obtained and checked to the Einstein field equations. In Sec.~\ref{sec-ec} the weak and strong energy conditions are studied. We study the main properties of geometry of the new rotating black hole spacetime, such as singularity, horizon and ergoregion in Sec.~\ref{sec-prop}. In Sec.~\ref{sec-hj} the equations of motion and the separation of the Hamilton-Jacobi equation are studied. In Sec.~\ref{sec-circular} general formalism for the circular orbits of the neutral test particle around the generic rotating regular black hole are shown. The epicyclic frequencies related to the oscillatory motion of a neutral test particle in the stable circular orbits around rotating Maxwellian black holes are studied in Sec.~\ref{sec-epicyclic}. Finally, we present some concluding remarks in Sec.~\ref{sec-conc}. Throughout the paper we use the geometric system of units $c=G=\hbar=1$ and a spacelike signature $(-,+,+,+)$.

\section{Newman-Janis algorithm}\label{sec-nj}

In the case of the generic spacetimes introduced in~\cite{Fan-Wan:2016:arxiv1610.02636:}, in order to obtain the rotating black hole solution we do not directly follow the standard Newman-Janis algorithm which was used in~\cite{New-Jan:1965:JMP:,Bam-Mod:2013:PhysLet:,Tos-Abd-Ahm-Stu:2014:PHYSR4:,Gho:2015:EPJC:}, however rather we follow the modified Newman-Janis algorithm~\cite{Azreg-Ainou2014PRD,Azreg-AinouEPJC:2015,Tos:Arxiv:2015} to avoid the number of appearing inadequate properties.\footnote{Transformation functions which are introduced at the last step of the Newman-Janis algorithm to turn into the Boyer-Lindquist coordinates from the Eddington-Finkelstein ones are not only $r$, but also $\theta$ dependent. Therefore, these coordinate transformations do not satisfy integrability condition. This creates several issues -- for details see~\cite{Azreg-Ainou2014PRD}.}

Let us consider the static, spherically symmetric line element in the Schwarzschild coordinates given in the form
\begin{eqnarray}\label{01}
ds^2=-f(r)dt^2+\frac{1}{f(r)}dr^2+h(r)(d\theta^2+\sin^2\theta d\phi^2).
\end{eqnarray}
At the first step, we turn the spacetime
metric~(\ref{01}) from the Schwarzschild coordinates ($t,r,\theta,\phi$)
to the Eddington-Finkelstein (EF) coordinates
($u,r,\theta,\phi$) by the coordinate transformations:
\begin{eqnarray}\label{2.1}
du=dt-\frac{dr}{f}\ .
\end{eqnarray}
In this way we obtain the spacetime metric in the form
\begin{eqnarray}\label{2.2}
ds^2=-fdu^2-2dudr+hd\theta^2+h\sin^2\theta d\phi^2\ .
\end{eqnarray}
Then the contravariant components
of the metric tensor in the advanced null
Eddington-Finkelstein (EF) coordinates can be
expressed in the form using the null tetrad,
\begin{eqnarray}\label{2.3}
g^{\mu\nu}=-l^\mu n^\nu-l^\nu n^\mu+m^\mu\bar{m}^\nu+m^\nu\bar{m}^\mu\ ,
\end{eqnarray}
where
\begin{eqnarray}\label{2.4}
&&l^\mu=\delta_r^\mu\ , \quad n^\mu=\delta_u^\mu-\frac{f}{2}\delta_r^\mu\ ,\nonumber\\
&&m^\mu=\frac{1}{\sqrt{2h}}\delta_\theta^\mu+ \frac{i}{\sqrt{2h}\sin\theta}\delta_\phi^\mu\ ,\nonumber\\ &&\bar{m}^\mu=\frac{1}{\sqrt{2h}}\delta_\theta^\mu- \frac{i}{\sqrt{2h}\sin\theta}\delta_\phi^\mu\ .
\end{eqnarray}
Vectors $\textbf{l}$ and $\textbf{n}$ are real, and the $\textbf{m}$ is the
complex vector, $\bar{\textbf{m}}$ vector is a
complex conjugate of the vector $\textbf{m}$.
They satisfy orthogonality
$l^\mu m_\mu=l^\mu \bar{m}_\mu=n^\mu m_\mu=n^\mu \bar{m}_\mu=0$,
isotropic $l^\mu l_\mu=n^\mu n_\mu=m^\mu m_\mu=\bar{m}^\mu \bar{m}_\mu=0$,
and normalization $l^\mu n_\mu=1$, $m^\mu\bar{m}_\mu=-1$ conditions.

Performing complex coordinate
transformations in the $u-r$ plane
\begin{eqnarray}\label{2.5}
u\rightarrow u-ia\cos\theta, \quad r\rightarrow r-ia\cos\theta ,
\end{eqnarray}
we can assume that as the result of
these transformations the metric
functions turn into a new form:
$f(r)\rightarrow F(r,a,\theta)$,
$h(r)\rightarrow \Sigma(r,a,\theta)$.
In the case $a=0$ new functions reduce
to initial forms. Null tetrads thus also take the form
\begin{eqnarray}\label{2.6}
&&l^\mu=\delta_r^\mu\ , \quad n^\mu=\delta_u^\mu-\frac{1}{2}F\delta_r^\mu\ , \nonumber\\ &&m^\mu=\frac{1}{\sqrt{2\Sigma}}\left[\delta_\theta^\mu+ia\sin\theta(\delta_u^\mu -\delta_r^\mu)+\frac{i}{\sin\theta}\delta_\phi^\mu\right]\ ,\nonumber\\ &&\bar{m}^\mu=\frac{1}{\sqrt{2\Sigma}}\left[\delta_\theta^\mu-ia\sin\theta (\delta_u^\mu-\delta_r^\mu)-\frac{i}{\sin\theta}\delta_\phi^\mu\right]\ .
\end{eqnarray}
Then we can rewrite the contravariant
nonzero components of the metric
tensor $g^{\mu\nu}$ by using~(\ref{2.3}) as
\begin{eqnarray} \label{2.7}
&&g^{uu}=\frac{a^2\sin^2\theta}{\Sigma}, \quad g^{ur}=-1-\frac{a^2\sin^2\theta}{\Sigma}, \quad g^{u\phi}=\frac{a}{\Sigma},\nonumber\\
&&g^{rr}=F+\frac{a^2\sin^2\theta}{\Sigma}, \quad
g^{r\phi}=-\frac{a}{\Sigma}, \quad g^{\theta\theta}=\frac{1}{\Sigma}, \nonumber\\&&g^{\phi\phi}=\frac{1}{\Sigma\sin^2\theta}.
\end{eqnarray}
The covariant nonzero components of the metric tensor read
\begin{eqnarray} \label{2.8}
&&g_{uu}=-F, \quad g_{ur}=-1, \quad g_{u\phi}=a\left(F-1\right)\sin^2\theta, \nonumber\\
&&g_{r\phi}=a\sin^2\theta, \quad g_{\theta\theta}=\Sigma, \nonumber\\ &&g_{\phi\phi}=\sin^2\theta\left[\Sigma+a^2\left(2-F\right)\sin^2\theta\right].
\end{eqnarray}
The last step of the Newman-Janis algorithm is
the turn back from the EF coordinates
to the Boyer-Lindquist (BL) coordinates by using the
following coordinate transformations:
\begin{eqnarray}\label{2.9}
du=dt+\lambda(r)dr\ , \quad d\phi=d\phi+\chi(r)dr\ .
\end{eqnarray}
The transformation functions
$\lambda(r)$ and $\chi(r)$ are
found due to the requirement that
all the nondiagonal components of
the metric tensor, except the
coefficient $g_{t\phi}$ ($g_{\phi t}$),
are equal to zero~\cite{Azreg-Ainou2014PRD,Azreg-AinouEPJC:2015}.
Thus,
\begin{eqnarray}\label{2.10}
\lambda(r)=-\frac{h(r)+a^2}{f(r)h(r)+a^2}\ , \quad \chi(r)=-\frac{a}{f(r)h(r)+a^2}
\end{eqnarray}
and
\begin{eqnarray}\label{2.12}
F(r,\theta)=\frac{fh+a^2\cos^2\theta}{h+a^2\cos^2\theta}.
\end{eqnarray}
Considering $h(r)=r^2$, and replacing (\ref{2.10}) and (\ref{2.12}) to (\ref{2.8}), we can write the rotating black hole spacetime metric in the BL coordinates
\begin{eqnarray}\label{rotating}
ds^2&&=-\frac{r^2f+a^2\cos^2\theta}{r^2+a^2\cos^2\theta}dt^2+ \frac{r^2+a^2\cos^2\theta}{r^2f+a^2}dr^2\nonumber\\
&&-2a\sin^2\theta\frac{r^2(1-f)}{r^2+a^2\cos^2\theta}d\phi dt+(r^2+a^2\cos^2\theta)d\theta^2\nonumber\\
&&+\sin^2\theta\left[r^2+a^2+a^2\sin^2\theta \frac{r^2(1-f)}{r^2+a^2\cos^2\theta}\right]d\phi^2\ .
\end{eqnarray}

\section{Generic rotating regular black hole solution of nonlinear electrodynamics}\label{sec-rbh}

By introducing the new notations we rewrite the spacetime metric (\ref{rotating}) in more compact, i.e., Kerr-like form as
\begin{eqnarray}\label{rotating2}
ds^2&=&-\left(1-\frac{2\rho r}{\Sigma}\right)dt^2+\frac{\Sigma}{\Delta}dr^2-2a\sin^2\theta\frac{2\rho r}{\Sigma}d\phi dt\nonumber\\
&&+\Sigma d\theta^2+\sin^2\theta\frac{(r^2+a^2)^2-a^2\Delta\sin^2\theta}{\Sigma}d\phi^2\ ,
\end{eqnarray}
or
\begin{eqnarray}\label{rotating3}
ds^2&=&-\frac{\Delta}{\Sigma}\left(dt-a\sin^2\theta d\phi\right)^2 +\frac{\Sigma}{\Delta}dr^2+\Sigma d\theta^2 \nonumber\\ &&+\frac{\sin^2\theta}{\Sigma}\left[(r^2+a^2)d\phi-adt\right]^2
\end{eqnarray}
where
\begin{eqnarray}\label{notations}
&&\Sigma=r^2+a^2\cos^2\theta, \qquad 2\rho=r(1-f),\nonumber\\
&&\Delta=r^2f+a^2=r^2-2\rho r+a^2\ .
\end{eqnarray}
Nonzero components of the Einstein tensor $G_{\mu\nu}$ read
\begin{eqnarray}\label{Einstein}
G_{tt}&=&\frac{2\left(r^4+a^2r^2-a^4\sin^2\theta\cos^2\theta\right)}{\Sigma^3}\rho' \nonumber\\ &&-\frac{4r^3}{\Sigma^3}\rho\rho'-\frac{a^2r\sin^2\theta }{\Sigma^2}\rho'',\nonumber\\
G_{rr}&=&-\frac{2r^2}{\Sigma\Delta}\rho',\nonumber\\
G_{t\phi}&=&\frac{2a\sin^2\theta(r^2+a^2)(a^2\cos^2\theta-r^2)}{\Sigma^3}\rho'\nonumber\\ &&+\frac{4ar^3\sin^2\theta}{\Sigma^3}\rho\rho'+\frac{a^2r\sin^2\theta (r^2+a^2)}{\Sigma^2}\rho'',\\ \nonumber
G_{\theta\theta}&=&-\frac{2a^2\cos^2\theta}{\Sigma}\rho'-r\rho'',\\ \nonumber
G_{\phi\phi}&=&-\frac{a^2\sin^2\theta(r^2+a^2)(a^2+(2r^2+a^2)\cos2\theta)}{\Sigma^3}\rho'\nonumber\\ &&-\frac{4a^2 r^3\sin^4\theta}{\Sigma^3}\rho\rho' -\frac{r\sin^2\theta (r^2+a^2)^2}{\Sigma^2}\rho''.\nonumber
\end{eqnarray}
The prime ($'$) stands for the derivative with respect to radial coordinate $r$. One can see from the Einstein tensor~(\ref{Einstein}) that in the case of constant mass function ($\rho=Const$) all components of the Einstein tensor vanish, i.e., spacetime metric (\ref{rotating2}) represents the vacuum solution of the Einstein field equation, namely, Kerr black hole solution. Now we use the Einstein equations $G_{\mu\nu}=8\pi T_{\mu\nu}$, where $T_{\mu\nu}$ is the energy-momentum tensor of the electromagnetic field that can be represented by the projection to a properly chosen tetrad of the vector and it can be represented by the expression
\begin{eqnarray}\label{set}
T^{(\mu)(\nu)}=e_\alpha^{(\mu)}e_\beta^{(\nu)}T^{\alpha\beta}.
\end{eqnarray}
The vector $e_\alpha^{(\mu)}$ is the orthonormal basis, here being the so-called Carter tetrad given by Eq.~(\ref{rotating3}) that can be written in the following form~\cite{CarterPR:1968}:
\begin{eqnarray}\label{basis}
&&e_t^{(\mu)}=\frac{1}{\sqrt{\Sigma\Delta}}\left(r^2+a^2,0,0,a\right), \quad
e_\theta^{(\mu)}=\frac{1}{\sqrt{\Sigma}}\left(0,0,1,0\right), \\
&&e_r^{(\mu)}=\sqrt{\frac{\Delta}{\Sigma}}\left(0,1,0,0\right), \quad
e_\phi^{(\mu)}=-\frac{1}{\sqrt{\Sigma}\sin\theta}\left(a\sin^2\theta,0,0,1\right).\nonumber
\end{eqnarray}
Now we determine related energy-momentum tensor $T_{\mu\nu}$ of the generic rotating regular solution. The components of the energy-momentum tensor $T^{(\mu)(\nu)}=(\epsilon,p_r,p_\theta,p_\phi)$ read
\begin{eqnarray}\label{set01}
&&8\pi\epsilon=-e_t^{(\mu)} e_t^{(\nu)} G_{\mu\nu}, \quad 8\pi p_r=g^{rr}G_{rr},\nonumber\\
&& 8\pi p_\theta=g^{\theta\theta}G_{\theta\theta}, \quad 8\pi p_\phi=-e_\phi^{(\mu)} e_\phi^{(\nu)}G_{\mu\nu}.
\end{eqnarray}
From the above Eq.~(\ref{set01}) we find the components of the stress-energy tensor
\begin{eqnarray}\label{set02}
\epsilon=-p_r=\frac{2\rho'r^2}{8\pi\Sigma^2}, \quad p_\theta= p_\phi= p_r-\frac{\rho''r+2\rho'}{8\pi\Sigma}.
\end{eqnarray}
Thus, we have shown that the spacetime metric~(\ref{rotating2}) is compatible with the Einstein field equations.

However, the Newman-Janis algorithm for generating rotating solutions is not always leading to true solutions of the whole set of field equations of the theory, i.e, the energy-momentum tensor of the rotating regular black hole solution generated by the Newman-Janis algorithm sometimes does not correspond to the nonlinear electrodynamics. Therefore, below we check validity of the resulting energy momentum tensor in the framework of the nonlinear electrodynamic field. It is known that the action of the regular black holes in general relativity coupled to the nonlinear electrodynamics reads
\begin{eqnarray}\label{action}
S=\frac{1}{16\pi}\int d^4x\sqrt{-g}\left(R-\mathscr{L}\right)
\end{eqnarray}
where $g$ is the determinant of the metric tensor $g_{\mu\nu}$, $R$ is the scalar curvature, and $\mathscr{L}$ represents the Lagrangian density of the nonlinear electrodynamic field that is the function of the nonlinear electrodynamic field strength $\mathscr{L}=\mathscr{L}\cal{(F)}$ with $\cal{F}=F_{\mu\nu}F^{\mu\nu}$. The equations of motion are derived from the action~(\ref{action}) in the form
\begin{eqnarray}
&&T_{\mu\nu}=2\left(\mathscr{L}_{\cal{F}}F_\mu^\alpha F_{\nu\alpha}- \frac{1}{4}g_{\mu\nu}\mathscr{L}\right),\label{eq-motion1}\\
&&\nabla_\mu\left(\mathscr{L}_{\cal{F}}F^{\mu\nu}\right)=0,\label{eq-motion2}
\end{eqnarray}
where $T_{\mu\nu}$ is the energy-momentum tensor of the nonlinear electrodynamics and $\mathscr{L}_{\cal{F}}=\frac{\partial\mathscr{L}}{\partial\cal{F}}$. The gauge field of the spherically symmetric regular black hole solution with the magnetic charge is $A_\mu=Q_{m}\cos\theta \delta_\mu^\phi$, i.e., only the last components survives. However, in the rotating spacetimes the gauge also changes and an extra component appears~\cite{Erbin:GRG:2015}:
\begin{eqnarray}\label{gauge}
A_\mu=-\frac{Q_{m}a\cos\theta}{\Sigma}\delta_\mu^t+ \frac{Q_{m}(r^2+a^2)\cos\theta}{\Sigma}\delta_\mu^\phi.
\end{eqnarray}
Thus, by calculating the covariant and contravariant electromagnetic field tensors we obtain
\begin{eqnarray}\label{squaredF}
&&{\cal{F}}=\\&&\frac{Q_{m}^2[a^4(3-\cos4\theta)+4(6a^2r^2+2r^4 +a^2(a^2-6r^2)\cos2\theta)]}{4\Sigma^4}\nonumber
\end{eqnarray}
In the case of $a=0$, we recover the field related to the spherically symmetric case ${\cal{F}}=2Q_{m}^2/r^4$~\cite{Fan-Wan:2016:arxiv1610.02636:}.

By inserting the Einstein tensors~(\ref{Einstein}) into the equations of motion~(\ref{eq-motion1}) ($G_{\mu\nu}=T_{\mu\nu}$), and solving them with respect to $\mathscr{L}$ and $\mathscr{L}_{\cal{F}}$, we obtain
\begin{widetext}
\begin{eqnarray}
&&\mathscr{L}=\frac{r^2[\left(15a^4-8a^2r^2+8r^4+4a^2(5a^2-2r^2)\cos2\theta+5 a^4\cos4\theta\right)\rho'+16a^2r\cos^2\theta\Sigma\rho'']}{2\Sigma^4},\label{lagrangian}\\
&&\mathscr{L}_{\cal{F}}=\frac{2(r^2-a^2\cos^2\theta)\rho'-r\Sigma\rho''}{2Q_{m}^2}.\label{lagrangian2}
\end{eqnarray}
\end{widetext}
In the non-rotating spacetime limit, $a=0$, we recover $\mathscr{L}=4\rho'/r^2$ and $\mathscr{L}_{\cal{F}}=r^2(2\rho'-r\rho'')/2Q_m^2$~\cite{Fan-Wan:2016:arxiv1610.02636:}. In the Maxwellian limit, namely, $\mathscr{L}={\cal{F}}$, from Eqs.~(\ref{squaredF}) and~(\ref{lagrangian}), by solving the differential equation, one obtains
\begin{eqnarray}\label{Max-limit}
\rho(r)=Const-\frac{Q_m^2}{2r}.
\end{eqnarray}
If we put the mass function~(\ref{Max-limit}) into the rotating spacetime metric~(\ref{rotating2}), we arrive at the Kerr-Newman solution and as it was pointed out in~\cite{Fan-Wan:2016:arxiv1610.02636:}, that in the case of $a=0$ we recover the Reissner-Nordstr\"{o}m one.

By choosing the mass function, one can obtain several rotating black hole solutions. In this paper we took the spherically symmetric regular black hole spacetimes in general relativity (GR) coupled to nonlinear electrodynamics, obtained by Fan and Wang~\cite{Fan-Wan:2016:arxiv1610.02636:}
\begin{eqnarray}\label{general}
\rho(r)=M+\frac{\alpha^{-1}q^3r^{\mu}}{(r^\nu+q^\nu)^{\mu/\nu}}.
\end{eqnarray}
Here $M$ denotes the pure Schwarzschild gravitational self-interaction mass.
The Arnowitt-Deser-Misner (ADM) mass of the spacetime is given by the relation~\cite{Fan-Wan:2016:arxiv1610.02636:}
\begin{eqnarray}\label{mass}
M_{ADM}=M+M_{em},
\end{eqnarray}
where $M_{em}$ is the electromagnetically induced gravitational mass and it reads
\begin{eqnarray}\label{mass}
M_{em}=\frac{q^3}{\alpha},
\end{eqnarray}
and is determined by the solution of the GR Einstein equations coupled to the nonlinear electrodynamics. The magnetic (electric) charge is expressed by the introduced parameters in the form~\cite{Fan-Wan:2016:arxiv1610.02636:}
\begin{eqnarray}\label{charge}
Q_{m}=\frac{q^2}{\sqrt{2\alpha}}.
\end{eqnarray}
The nonlinear electrodynamics is governed by the dimensionless constant parameter $\mu$ characterizing the degree of nonlinearity, and the parameter $\alpha$ with the dimension of length squared, governing strength of the nonlinear effects. Parameters $q$ and $\nu$ are free parameters governing the magnetic (electric) charge, and character of its field in concrete solutions-- for details see~\cite{Fan-Wan:2016:arxiv1610.02636:}. In this framework several classes of the known regular black hole solutions can be obtained. For example, when $\nu=2$, the function~(\ref{general}) represents the mass function of the Bardeen-like black hole spacetime
\begin{eqnarray}\label{bardeen}
\rho(r)=M+\frac{\alpha^{-1}q^3r^{\mu}}{(r^2+q^2)^{\mu/2}}.
\end{eqnarray}
In the case of $\nu=\mu$, it corresponds to the Hayward-like black holes
\begin{eqnarray}\label{hayward}
\rho(r)=M+\frac{\alpha^{-1}q^3r^{\mu}}{r^\mu+q^\mu}.
\end{eqnarray}
For $\nu=1$, new class of generic black hole solutions is obtained
\begin{eqnarray}\label{new1}
\rho(r)=M+\frac{\alpha^{-1}q^3r^{\mu}}{(r+q)^{\mu}}.
\end{eqnarray}
that correspond to the Maxwell equations in the weak-field limit of the nonlinear electrodynamics.

By replacing the expression~(\ref{mass}) and turning into the dimensionless coordinates: $t/M_{em}\rightarrow t$, $r/M_{em}\rightarrow r$, and dimensionless parameters: $M/M_{em}\rightarrow M$, $q/M_{em}\rightarrow q$, (for the rotating solution we similarly introduce also the dimensionless spin parameter by $a/M_{em}\rightarrow a$) we rewrite the mass function for the Bardeen-like black holes~(\ref{bardeen})
\begin{eqnarray}\label{bardeen1}
\rho(r)=M+\frac{r^{\mu}}{(r^2+q^2)^{\mu/2}},
\end{eqnarray}
for the Hayward-like black holes~(\ref{hayward})
\begin{eqnarray}\label{hayward1}
\rho(r)=M+\frac{r^{\mu}}{r^\mu+q^\mu},
\end{eqnarray}
for the new class of generic black holes (\ref{new1}) in the form
\begin{eqnarray}\label{new2}
\rho(r)=M+\frac{r^{\mu}}{(r+q)^{\mu}}.
\end{eqnarray}
The dimensionless specific charge parameter reads
\begin{eqnarray}
q=\frac{2Q_{m}^2}{M_{em}^2}.
\end{eqnarray}
Note that we have to be careful in order to obtain the Schwarzschild (Kerr) limit for the case of $q\rightarrow0$. Due to the definition of $M_{em}$, for $q\rightarrow0$, there is also $\alpha\rightarrow0$ in order to keep $M_{em}=Const\neq0$. Then $q=(\alpha M_{em})^{1/3}$ and the charge reads $Q_{m}=2^{1/2}M_{em}^{2/3}\alpha^{1/6}$, i.e., the charge vanishes when $q\rightarrow0$, while the mass parameter is fixed.

Note that existence of the Schwarzschild mass $M$ implies for all the spacetimes mentioned above existence of an unavoidable spacelike curvature singularity. In the case of absence of the gravitational mass ($M=0$), these spacetimes are regular. We address these singularity problems in the next section.

In the following, we concentrate on discussion of the spacetime properties and its implications.

\section{Energy conditions}\label{sec-ec}

In~\cite{Fan-Wan:2016:arxiv1610.02636:} it is mentioned that the strong energy condition (SEC) is violated in the non-rotating case, while the weak energy condition (WEC) is satisfied. Now we check these energy conditions in the case of the generic rotating regular black hole. In order to check the energy conditions we turn to the diagonal stress-energy tensor $T^{(a)(b)}=diag(T^{(0)(0)}, T^{(1)(1)}, T^{(2)(2)}, T^{(3)(3)})$ by using the following orthonormal tetrads which correspond to the standard locally non-rotating frame (LNRF)~\cite{BardeenApJ:1972}:
\begin{eqnarray} \label{tetrad}
e_\mu^{(a)}=\left(
\begin{array}{c c c c}
\sqrt{\mp(g_{tt}-\Omega g_{t\phi})} & 0 & 0 & 0 \\
0 & \sqrt{\pm g_{rr}} & 0 & 0 \\
0 & 0 & \sqrt{g_{\theta\theta}} & 0 \\
\Omega\sqrt{g_{t\phi}} & 0 & 0 & \sqrt{g_{\phi\phi}} \\
\end{array} \right).
\end{eqnarray}
where $\Omega=g_{t\phi}/g_{\phi\phi}$. The energy-momentum tensor is turned to diagonal form by the relation $T^{(a)(b)}=e_\mu^{(a)}e_\nu^{(b)}G^{\mu\nu}/8\pi$. The signature of (\ref{tetrad}) reflects the considered region. We claim that the
spacetime metric~(\ref{rotating2}) has two horizons: inner (Cauchy) and outer (event) horizons, in the case of black hole. If the considered region is located outside the event horizon or inside the inner horizon we choose the signature of the components of the orthonormal tetrads $e_0^{(0)}$ and $e_1^{(1)}$ as $(-,+)$, respectively. In the case of the region under consideration is located between these two horizons we take the signature of $e_0^{(0)}$ and $e_1^{(1)}$ as $(+,-)$, respectively. Now we are going to check the behaviour of the energy-momentum tensor near origin, i.e. inside the inner horizon. We do not show the full expression of components of the energy-momentum tensor because of their cumbersome form. For simplicity we consider the "poles" $\theta=0,\pi$. Then, the components of the energy-momentum tensor inside the inner horizon and outside the event horizon take the form
\begin{figure*}[th]
\begin{center}
\includegraphics[width=0.32\linewidth]{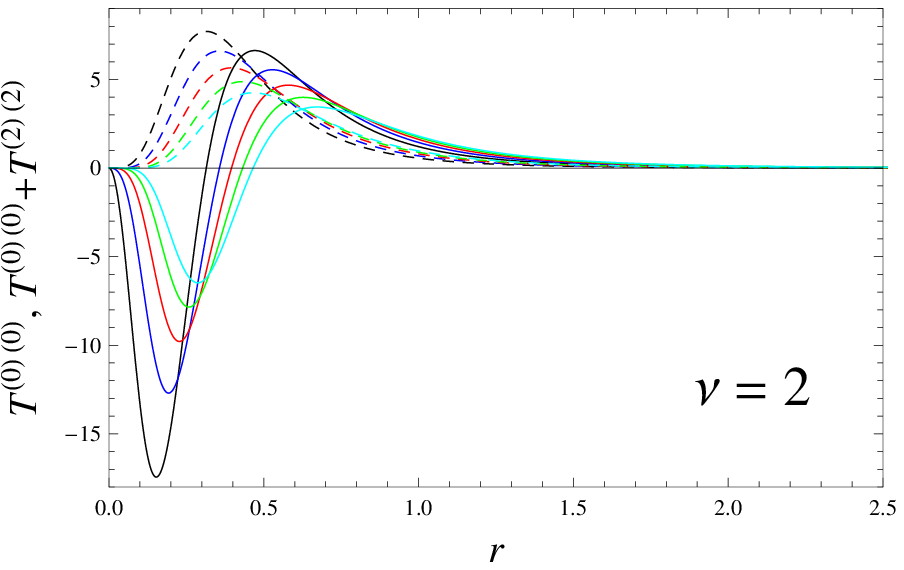}
\includegraphics[width=0.32\linewidth]{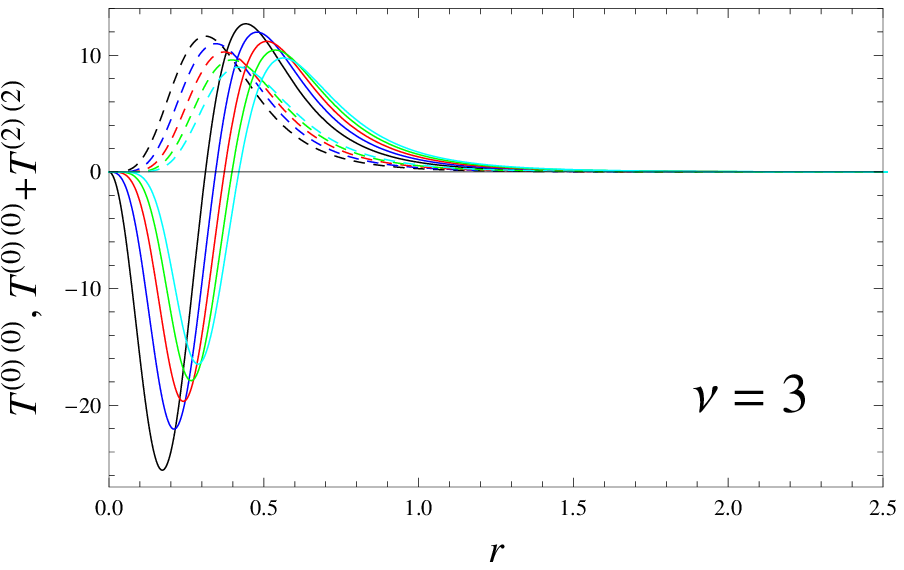}
\includegraphics[width=0.32\linewidth]{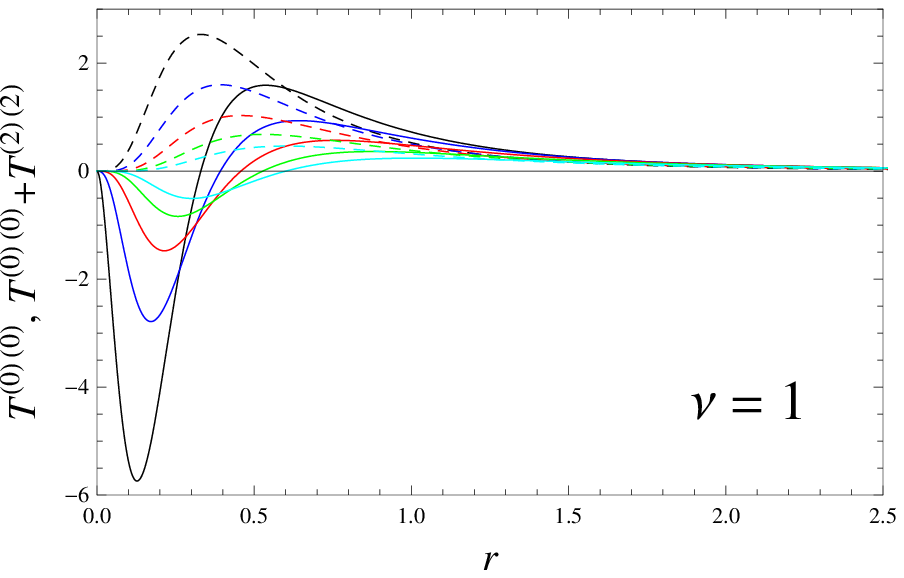}
\end{center}
\caption{\label{fig-wec} Radial profiles of the energy-momentum tensor in the LNRF. Plot of $T^{(0)(0)}$ (dashed) and $T^{(0)(0)}+T^{(2)(2)}$ (solid) for rotating (from left to right) Bardeen-like, Hayward-like and new type of black holes with $q=0.4$, $a=0.3$, and $\cos^2\theta=1$. Where $\mu=3$ -- black, $\mu=4$ -- blue,  $\mu=5$ -- red,  $\mu=6$ -- green, and $\mu=7$ -- cyan curves. }
\end{figure*}
\begin{figure*}[th]
\begin{center}
\includegraphics[width=0.32\linewidth]{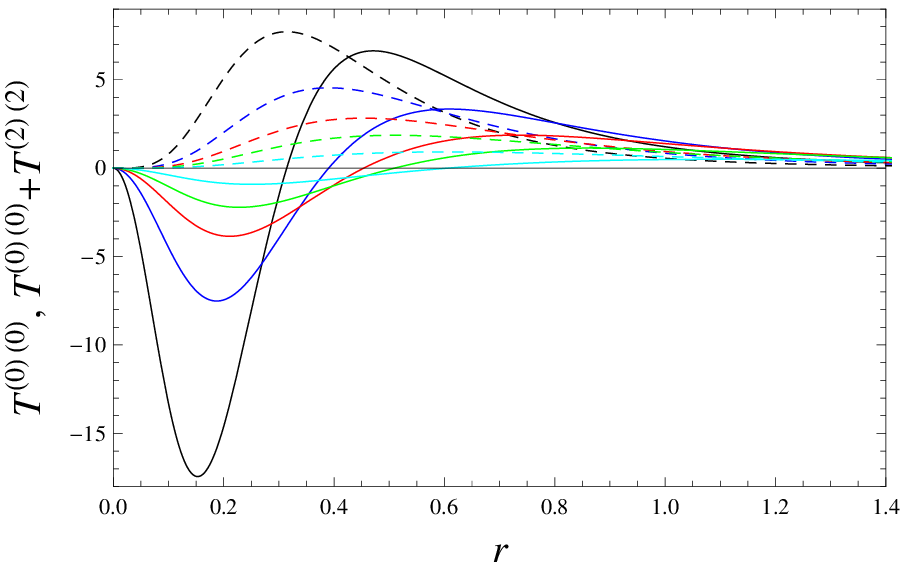}
\includegraphics[width=0.32\linewidth]{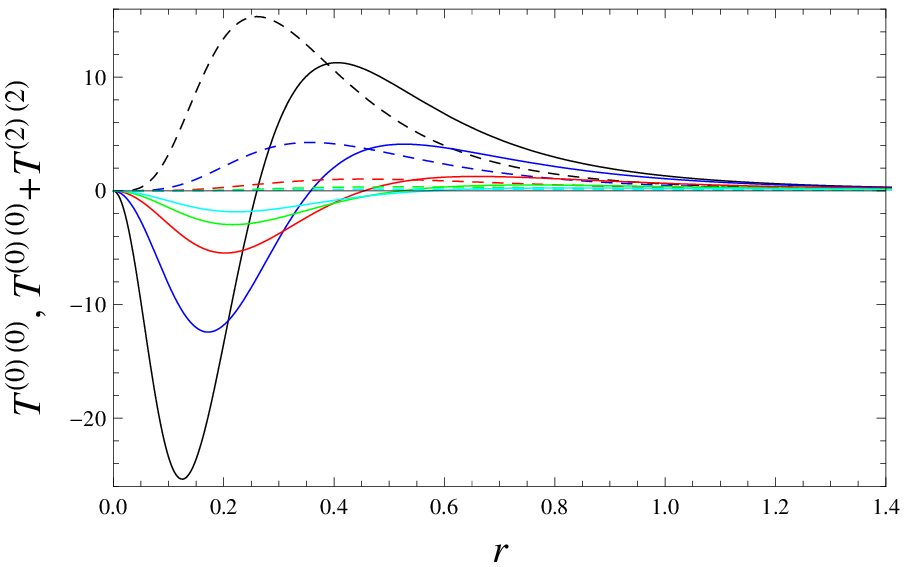}
\includegraphics[width=0.32\linewidth]{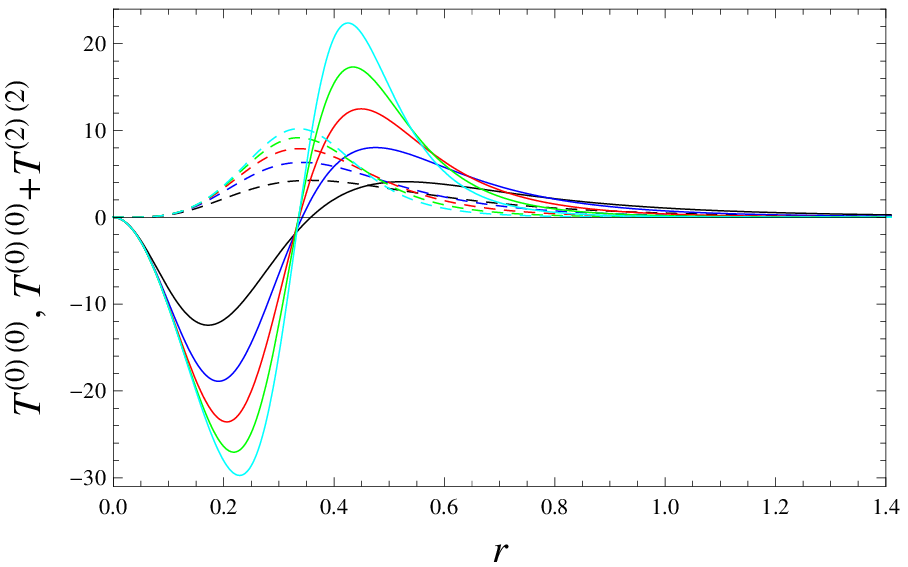}
\end{center}
\caption{\label{fig-ec} Radial profiles of the energy-momentum tensor in the LNRF. Plot of $T^{(0)(0)}$ (dashed) and $T^{(0)(0)}+T^{(2)(2)}$ (solid) for rotating regular Bardeen-like black holes for the values of the parameters. Left panel: $\mu=3$, $\nu=2$, $a=0.3$, and $\cos^2\theta=1$ for the values of the charge parameter: $q=0.4$ -- black, $q=0.6$ -- blue,  $q=0.8$ -- red,  $q=1.0$ -- green, and $q=1.4$ -- cyan curves. Middle panel: $\mu=3$, $\nu=2$, $q=0.4$, and $\cos^2\theta=1$ for the values of the rotation parameter: $a=0.2$ -- black, $a=0.4$ -- blue,  $a=0.7$ -- red,  $a=1.0$ -- green, and $a=1.3$ -- cyan curves. Right panel: $\mu=3$, $a=0.4$, $q=0.4$, and $\cos^2\theta=1$ for the values of $\nu$: $\nu=2$ -- black, $\nu=3$ -- blue,  $\nu=4$ -- red,  $\nu=5$ -- green, and $\nu=6$ -- cyan curves.}
\end{figure*}
\begin{eqnarray}\label{emtensor}
&&T^{(0)(0)}=\frac{2r^2\rho'}{(r^2+a^2)^2}=-T^{(1)(1)}\ ,\nonumber\\
&&T^{(2)(2)}=-\frac{r\rho''}{r^2+a^2}-\frac{2a^2\rho'}{(r^2+a^2)^2}=T^{(3)(3)}\ ,\\
&&T^{(0)(0)}+T^{(2)(2)}= \frac{2(r^2-a^2)\rho'}{(r^2+a^2)^2}-\frac{r\rho''}{r^2+a^2}\nonumber\\
&&=T^{(0)(0)}+T^{(3)(3)}\ ,\nonumber
\end{eqnarray}
One can see from (\ref{emtensor}) that the existence of the Schwarzschild mass does not effect the energy conditions of these black holes, since only derivatives of the $\rho$~\footnote{where $\rho=M+\frac{r^\mu}{(r^\nu+q^\nu)^{\mu/\nu}}$. Therefore, pure gravitational mass $M$ plays no role in energy conditions.} with respect to radial coordinate appear in energy-momentum tensor.
According to the WEC, $T_{\mu\nu}u^\mu u^\nu\geq0$ where $u^\mu$ is the generic timelike vector, or $T^{(0)(0)}\geq0$ and $T^{(0)(0)}+T^{(i)(i)}\geq0$, where $i=1,2,3$. One can see from Fig.~\ref{fig-wec} that near the origin (inside the inner horizon) the WEC is violated in the case of $a\neq0$. With increasing the value of the parameter $\mu$ that characterizes the degree of the nonlinearity of the electromagnetic field, a depth of the violation of the WEC decreases however, even for the large values of $\mu$ violation of the WEC does not vanish. Since the behaviours of the all class of regular spacetimes (Bardeen-like, Hayward-like and new class) relative to the parameters are the same, in Fig.~\ref{fig-ec} we present the behaviours of the spacetimes for different values of the spacetime parameters only for the Bardeen-like spacetime. From Fig.~\ref{fig-ec} one can see that with increasing the values of the rotation and parameters the depth of the violation of the WEC decreases, however, an increase in the value of the nonlinear electromagnetic field parameter $\nu$ the depth of the violation increases. Since the strong energy condition (SEC) requires the condition $(T_{\mu\nu}-Tg_{\mu\nu}/2)u^\mu u^\nu\geq0$, or $\sum_{a=0}^{4} T^{(a)(a)}\geq0$, violation of the WEC guarantees the violation of the SEC too.

\section{Properties of the rotating regular black hole solution}\label{sec-prop}

\subsection{Curvature singularity}

It is well known that in general relativity (GR) there are two types of singularities which are the points where the spacetime metric coefficients tend to infinity: curvature singularity which is often called a physical singularity by the reason of that it cannot be removed by coordinate transformations, while the second one is the coordinate singularity (event horizon) which can be considered as a mathematical singularity due to the possibility of elimination of it by introducing appropriate coordinate system.

Let us study first the physical singular points of the new rotating regular black hole~(\ref{rotating2}) by studying its curvature invariants, such as curvature scalar, Ricci square, and Kretschmann invariant. Naturally, we consider the case of the zero Schwarzschild mass ($M=0$). First, we approach the center ($r=0$) outside the equatorial plane:
\begin{eqnarray}\label{sing0}
&&\lim_{\theta\rightarrow [0,\pi]}\left(\lim_{r\rightarrow0}R\right)=\lim_{r\rightarrow 0}\left(\lim_{\theta\rightarrow\neq\pi/2}R\right)=0,\nonumber\\
&&\lim_{\theta\rightarrow [0,\pi]}\left(\lim_{r\rightarrow0} R_{\mu\nu}R^{\mu\nu}\right)=\lim_{r\rightarrow 0}\left(\lim_{\theta\rightarrow\neq\pi/2} R_{\mu\nu}R^{\mu\nu}\right)=0,\nonumber\\
&&\lim_{\theta\rightarrow [0,\pi]}\left(\lim_{r\rightarrow0} R_{\mu\nu\rho\sigma}R^{\mu\nu\rho\sigma}\right)\\
&&=\lim_{r\rightarrow 0}\left(\lim_{\theta\rightarrow\neq\pi/2} R_{\mu\nu\rho\sigma}R^{\mu\nu\rho\sigma}\right)=0.\nonumber
\end{eqnarray}
Second, we approach the center in the equatorial plane:
\begin{eqnarray}\label{sing1}
&&\lim_{r\rightarrow 0}\left(\lim_{\theta\rightarrow\pi/2}R\right)= \frac{24}{\alpha}r^{\mu-3}=\frac{24}{\alpha}\lim_{r\rightarrow 0}r^{\mu-3},\\
&&\lim_{r\rightarrow 0}\left(\lim_{\theta\rightarrow\pi/2}R_{\mu\nu}R^{\mu\nu} \right)=\frac{144}{\alpha}r^{2\mu-6}=\frac{144}{\alpha}\lim_{r\rightarrow 0}r^{2\mu-6},\nonumber\\
&&\lim_{r\rightarrow 0}\left(\lim_{\theta\rightarrow\pi/2} R_{\mu\nu\rho\sigma}R^{\mu\nu\rho\sigma}\right)=\frac{96}{\alpha}r^{2\mu-6}=\frac{96}{\alpha}\lim_{r\rightarrow 0}r^{2\mu-6}.\nonumber
\end{eqnarray}
It follows from (\ref{sing1}) that the rotating black hole solution (\ref{rotating2}) with zero Schwarzschild mass is regular everywhere only if $\mu\geq3$. One can see from the curvature invariants~(\ref{sing0}) and (\ref{sing1}) that in the case of $\mu=3$ the values of these invariants are dependent on the direction we approach toward the center and at the center there is a "de Sitter" -like core of the nonlinear electrodynamic source. For $\mu>3$, the values of these invariants are independent of the way we approach the center.

\subsection{Event horizons and separation of black hole and no-horizon spacetimes in the parameter space}

In the stationary case, the event horizon coincides with the outermost apparent horizon. Its location is given by the following second order partial differential equation~\cite{Schoen:1983}:
\begin{eqnarray}\label{extrinsic}
\nabla_\mu n^\mu+\gamma_{\mu\nu}K^{\mu\nu}=0\ ,
\end{eqnarray}
where $n_\mu=(-\alpha,0,0,0)$ is the timelike four-vector which is normal to the hypersurface at each event in the spacetime. From the normalization condition on timelike four-vectors, $n_\mu n^\mu=-1$, we find that
\begin{eqnarray}
\alpha=\frac{1}{\sqrt{-g^{tt}}}\ ,
\end{eqnarray}
$\gamma_{\mu\nu}$ is given by the relation
\begin{eqnarray}
\gamma_{\mu\nu}=g_{\mu\nu}+n_\mu n_\nu\ ,
\end{eqnarray}
$K_{\mu\nu}$ is the extrinsic curvature which is given by the relation~\cite{Rezzolla:book}
\begin{eqnarray}
K_{\mu\nu}=-\nabla_\mu n_\nu-n_\mu n^\sigma \nabla_\sigma n_\nu\ .
\end{eqnarray}
Simplifying Eq.~(\ref{extrinsic}) we derive the relation
\begin{eqnarray}\label{final}
g^{\mu\nu}+n^\mu n^\nu=0\ ,
\end{eqnarray}
From Eq.~(\ref{final}) one can easily obtain that the horizon of the spacetime metric~(\ref{rotating3}) is defined by the equations
\begin{eqnarray}\label{final2}
g^{rr}=0\ , \quad (g^{t\phi})^2-g^{tt}g^{\phi\phi}=0\ .
\end{eqnarray}
In~(\ref{final2}) the latter equation is equivalent to the first one. Thus, the coordinate singularity giving the event horizon of the black hole is determined by the relation $g^{rr}=0$, i.e.,
\begin{eqnarray}\label{hor}
\Delta=r_+^2-2\rho_+ r_++a^2=0\ ,
\end{eqnarray}
\begin{figure*}[th]
\begin{center}
\includegraphics[width=0.32\linewidth]{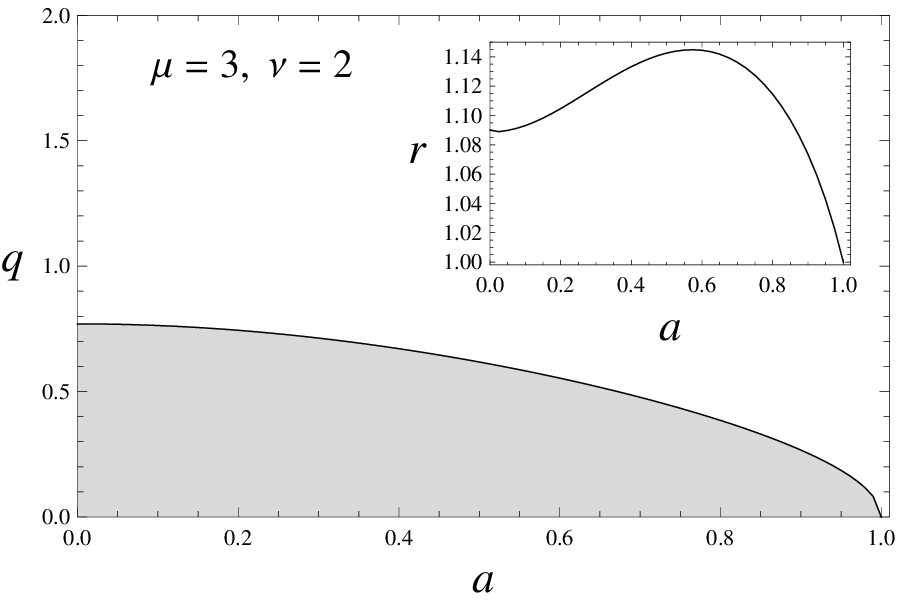}
\includegraphics[width=0.32\linewidth]{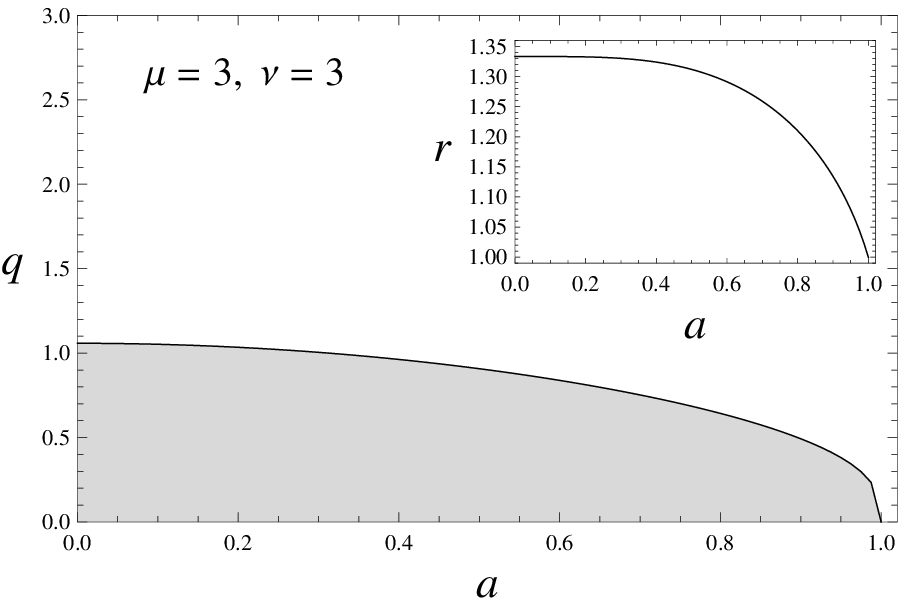}
\includegraphics[width=0.32\linewidth]{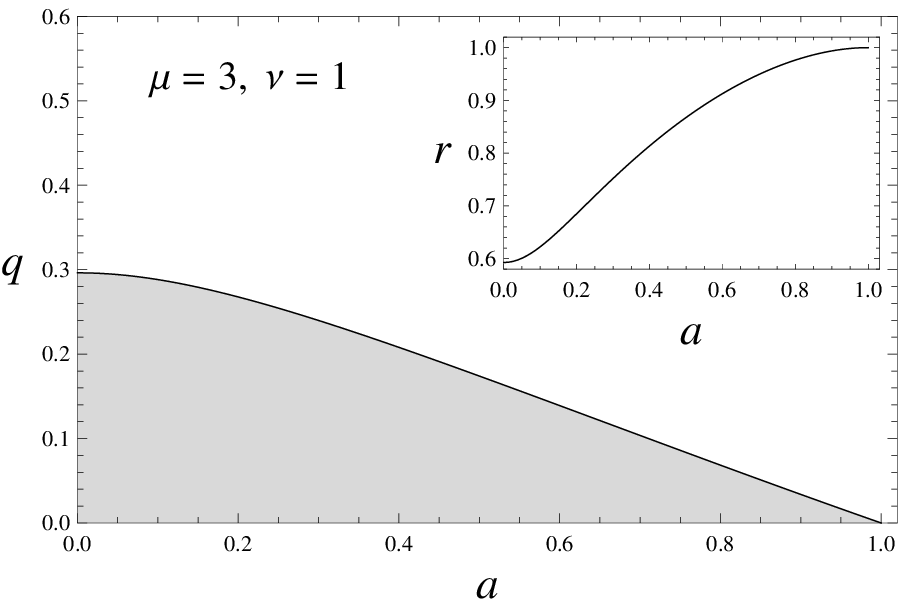}
\includegraphics[width=0.32\linewidth]{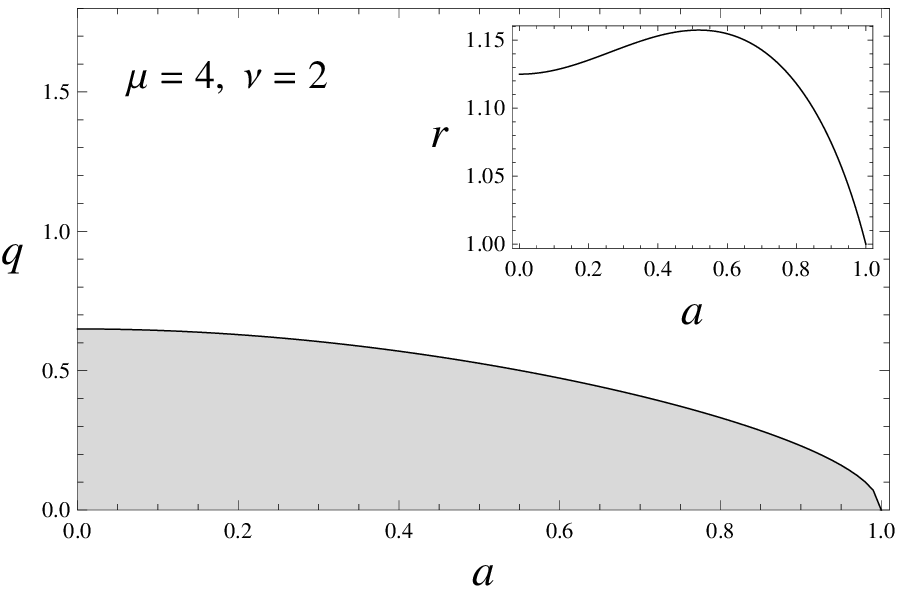}
\includegraphics[width=0.32\linewidth]{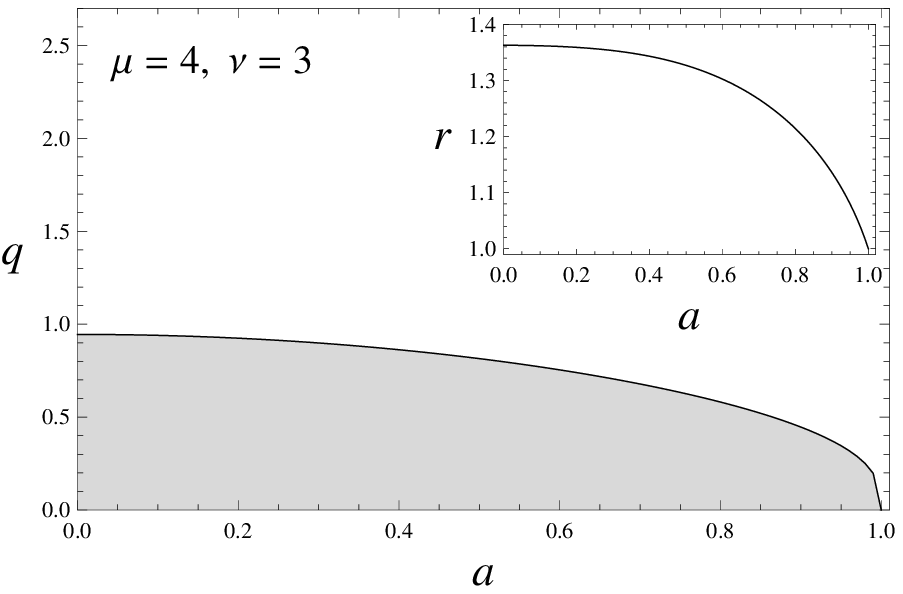}
\includegraphics[width=0.32\linewidth]{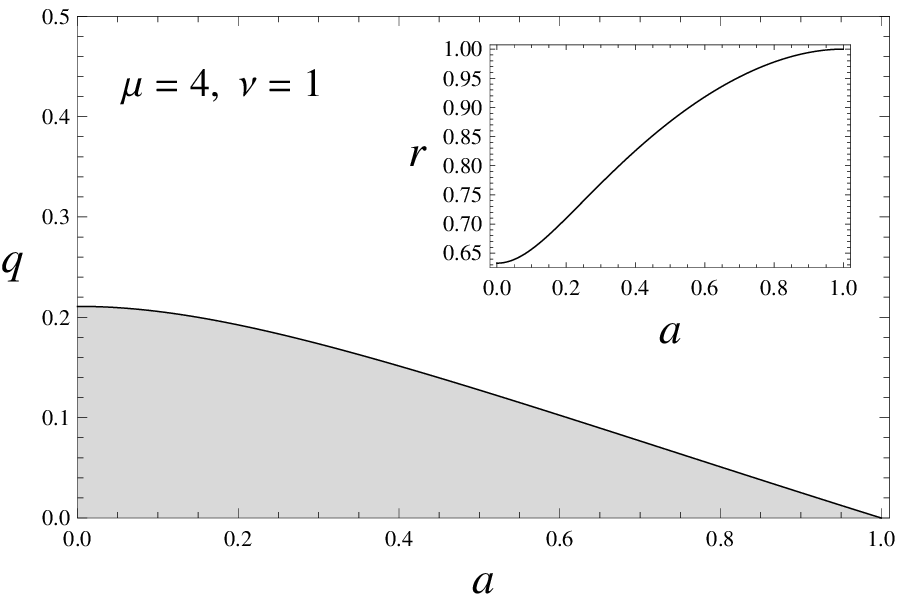}
\includegraphics[width=0.32\linewidth]{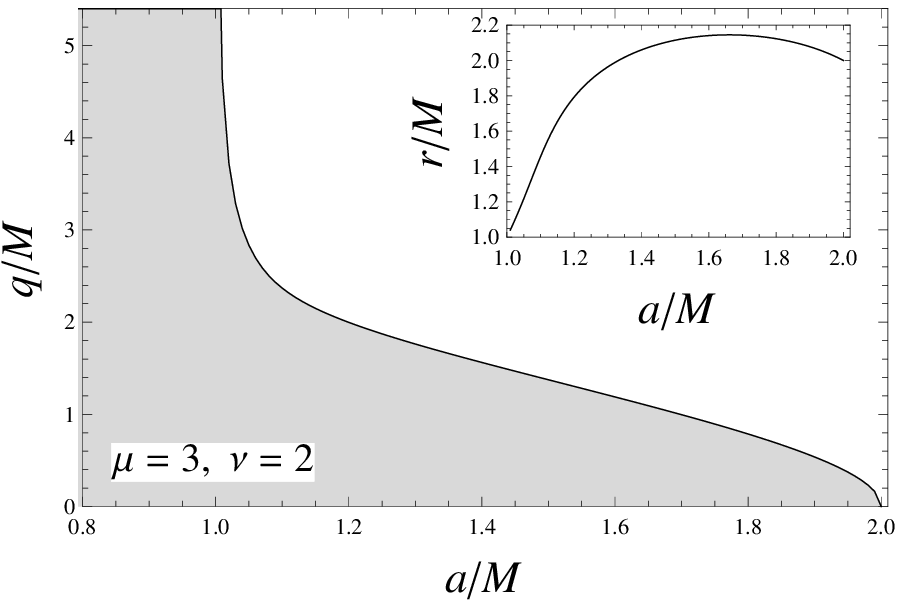}
\includegraphics[width=0.32\linewidth]{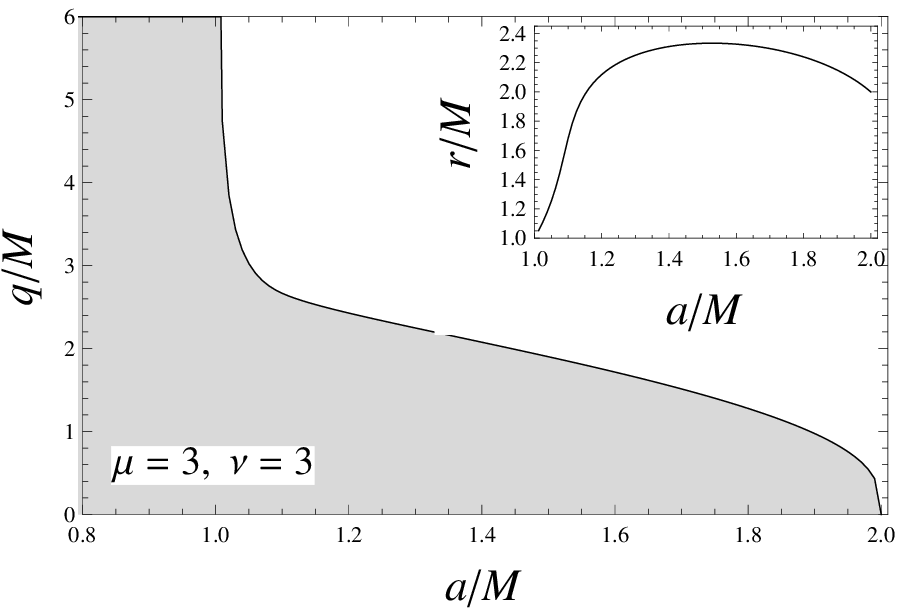}
\includegraphics[width=0.32\linewidth]{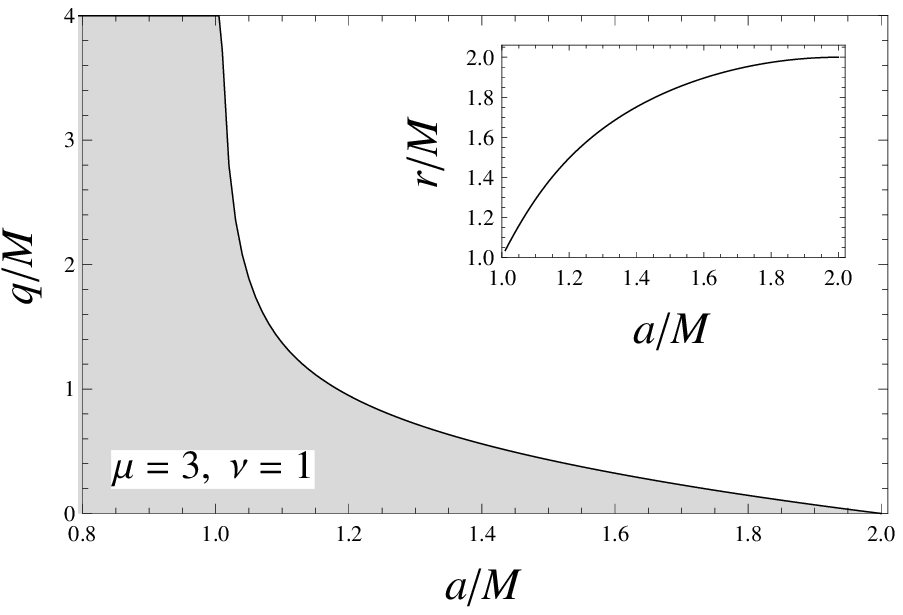}
\end{center}
\caption{\label{fig-hor} Separation of the black hole and no-horizon (naked singularity) spacetimes $q-a$ parameter space. \textit{\textbf{Top panel:}} (from left to right) rotating regular ($M=0$) Bardeen-like, Hayward-like and new type of regular black holes with $\mu=3$. In the $q-a$ plots the shaded regions represent the black holes with two horizons, while the white regions represent the no-horizon spacetimes. The border of the shaded and white regions represents the extremal regular black holes. $r-a$ plots represent the location of the horizon of the extremal rotating regular black holes. \textit{\textbf{Middle panel:}} The same as the top panel but for the case of $\mu=4$. \textit{\textbf{Bottom panel:}} the same as top panel but for the non-zero Schwarzschild mass $M\neq0$. Unlike the case of the rotating regular black holes (top panel), in $q/M-a/M$ plots white regions represent the naked singularity spacetime regions.}
\end{figure*}
or $r_+^2f(r_+)+a^2=0$. It is well known that the spacetime~(\ref{rotating2}) can represent several scenarios, such as regular black hole, singular black hole, no-horizon and naked singularity spacetimes. Regular black hole and no-horizon spacetimes are represented by the same line elements, while singular black hole and naked singularity spacetimes are represented by the different
alternate spacetimes. The black hole and no-horizon or naked singularity spacetimes are separated by the extremal horizon of the black hole. The extremality condition of the black hole horizon is defined by the relation
\begin{eqnarray}\label{extreme}
\Delta=0, \qquad \Delta'=0.
\end{eqnarray}
For the current black hole
\begin{eqnarray}\label{extreme2}
r(\rho-r\rho')-a^2=0,
\end{eqnarray}
or
\begin{eqnarray}\label{extreme3}
r\left[M+\frac{r^\mu(r^\nu-q^\nu(\mu-1))}{(r^\nu+q^\nu)^{\mu/\nu+1}}\right]-a^2=0.
\end{eqnarray}
In Fig.~\ref{fig-hor} the conditions of the black hole and the no-horizon spacetimes in terms of the values of the characteristic parameters of the spacetime metric are given for the generic rotating regular black hole with zero Schwarzschild mass ($M=0$) (top panel) and the spacelike singular black hole with the non-zero gravitational mass $M$ (bottom panel) for the case $\mu=3$ and some special values of the parameter $\nu$. Moreover, as in the case of the Kerr, Kerr-Newman or other rotating regular black holes~\cite{Tos-Abd-Ahm-Stu:2014:PHYSR4:}, with increasing value of the specific magnetic (electric) charge parameter, the radius of the black hole decreases. Interestingly, irrespective of the value of the magnetic (electric) charge parameter, all classes (Bardeen-like, Hayward-like and new type) of generic rotating black holes with the non-zero gravitational mass $M$ always have non-vanishing two horizons for the value of the rotation parameter $a/M<1$, i.e., rotating singular black hole with rotation parameter $a/M<1$, naked singularity does not exist regardless the value of the specific charge parameter. In the case of the rotating regular black holes ($M=0$), existence of the horizons is strongly dependent on the charge parameter. As in the case of the rotating Hayward~\cite{Ami-Gho:2015:JHEP:}, Bardeen~\cite{GhoshEPJC:2015} and ABG~\cite{Tos-Abd-Ahm-Stu:2014:PHYSR4:} non-singular spacetimes, in the case of the current rotating regular black holes, there are the upper limits on the values of the specific charge and rotation parameters in order for the spacetimes to represent black holes. Let us denote these limiting values as a critical values, $q_{cr}$ and $a_{cr}$, which correspond to the border of the black hole (shaded) and no-horizon (white) regions in top panel of Fig.~\ref{fig-hor}: at these critical values of the charge parameter black holes have the minimum horizon $r_{min}$. For the greater values of the charge parameter or rotation parameter than these critical values, $q>q_{cr}$ or $a>a_{cr}$ horizons vanish and the line element represent the no-horizon spacetimes. One can see from Fig.~\ref{fig-hor} that the rotating regular black holes can have smaller value of the specific charge parameter than the non-rotating one. With increasing the rotation parameter $a$, a possible value of the specific charge parameter $q_{cr}$ decreases and when rotation parameter reaches the its maximum value, $a=1$, in the case of the black hole, the charge parameter vanishes. In Fig.~\ref{fig-hor} it has been shown that among these three classes of regular black holes, the rotating Hayward-like regular black holes have the greatest critical value of the specific charge, $q_{cr}\approx 1.06$, while, the new type of rotating regular black holes have the smallest value of the specific charge, $q_{cr}\approx 0.3$. Moreover, with increasing the value of $\mu$ possible values of the specific charge and rotation parameters decrease.

The horizon area of the generic rotating black hole takes the standard form
\begin{eqnarray}\label{hor-area}
A_+=\int_0^{2\pi}d\phi\int_{0}^{\pi}d\theta\sqrt{g_{\theta\theta}g_{\phi\phi}}= 4\pi (r_+^2+a^2)\ .
\end{eqnarray}

\subsection{Ergosphere}

The static limit of the black hole is the one in which the timelike Killing vector $k^\mu=(1,0,0,0)$ becomes null. The static limit is determined by the relation
\begin{eqnarray}\label{sl}
g_{tt}(r_{sl})=r_{sl}^2-2\rho(r_{sl})r_{sl}+a^2\cos^2\theta=0.
\end{eqnarray}
\begin{figure*}[t]
\begin{center}
\includegraphics[width=0.32\linewidth]{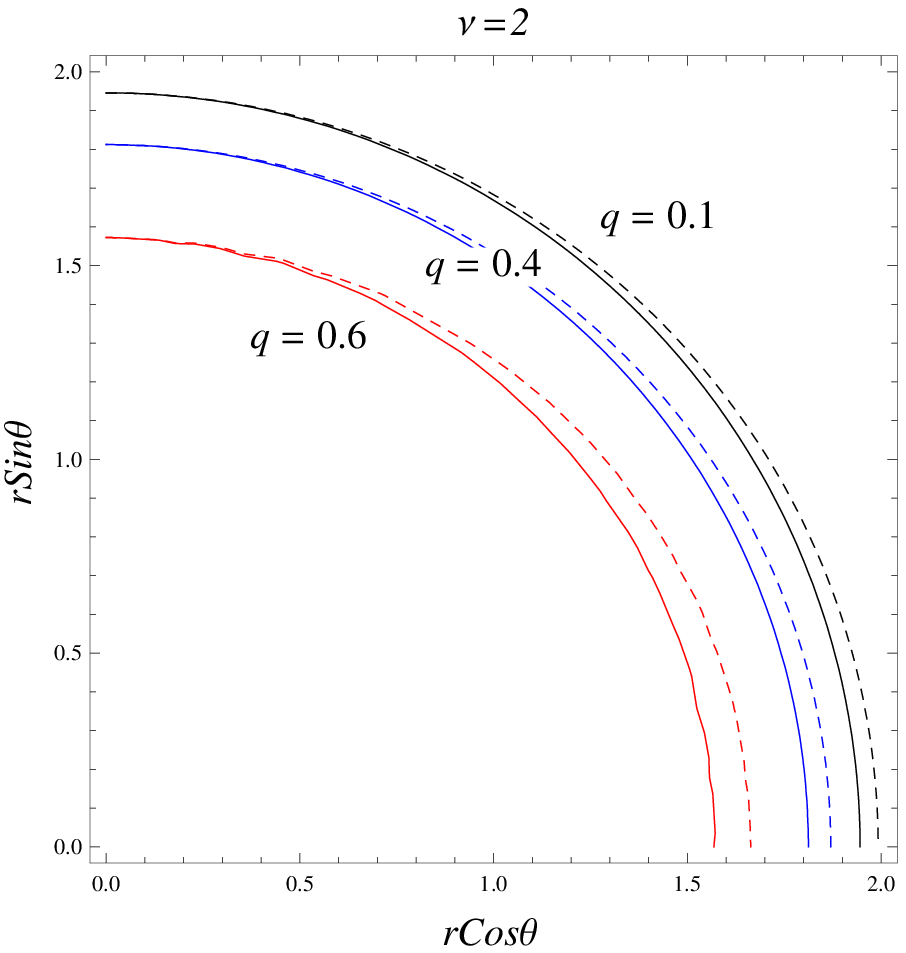}
\includegraphics[width=0.32\linewidth]{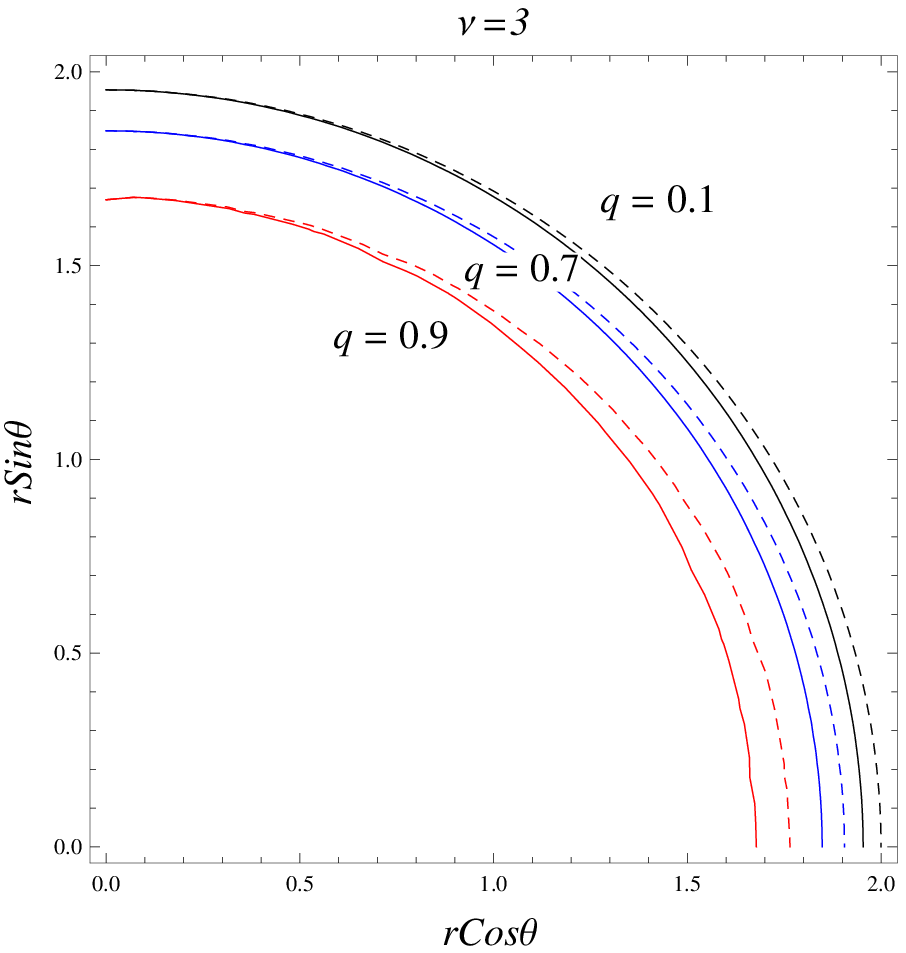}
\includegraphics[width=0.32\linewidth]{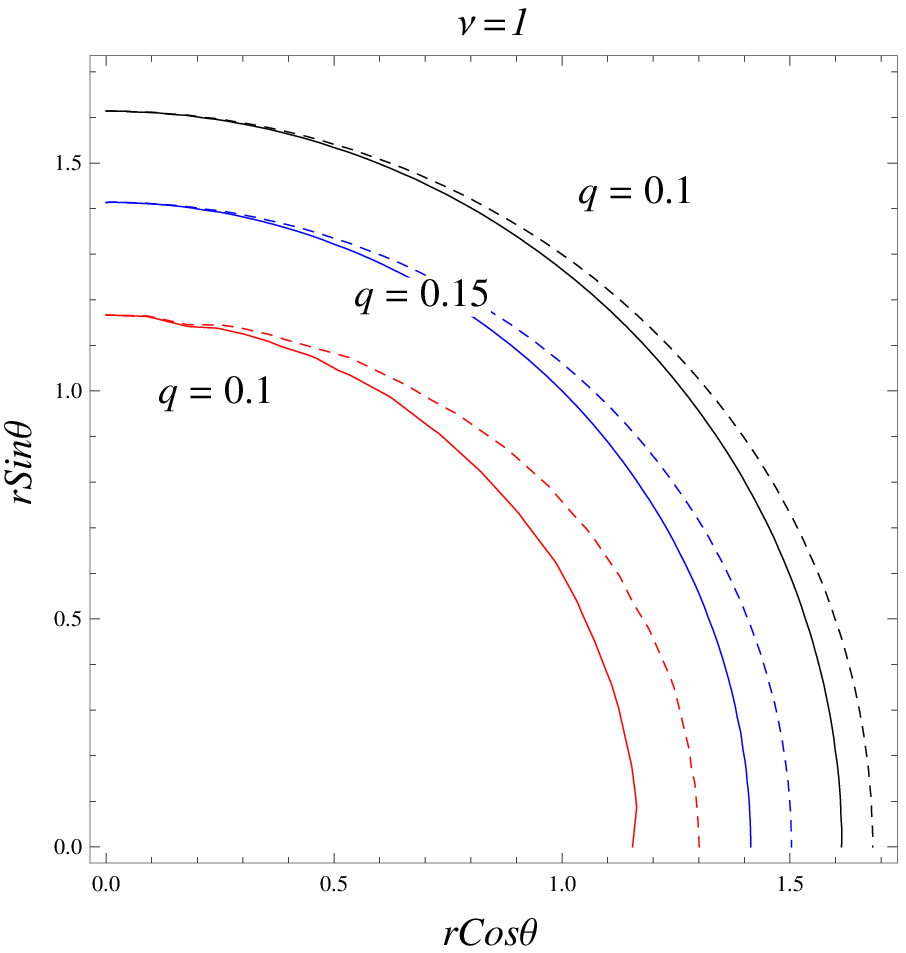}
\end{center}
\caption{\label{fig-ergosphere} Outer horizon (solid curves) and static limit (dashed curves) giving boundary of the ergosphere of the (from left to right) Bardeen-like, Hayward-like, and new type of rotating regular black holes are given for the values of the parameters $\mu=3$, $a=0.3$, and $q$ as depicted in the figures.}
\end{figure*}
One can see from (\ref{hor}) and (\ref{sl}) that at the poles $\theta=0,\pi$ the event horizon and static limit surface coincide ($r_+=r_{sl}$). The region between event horizon and static limit surface is called ergosphere or ergoregion. In Fig.~\ref{fig-ergosphere} the ergoregion of the generic rotating Bardeen-like, Hayward-like and new type of regular black hole has been plotted. One can see from Fig.~\ref{fig-ergosphere} with increasing the value of the charge parameter $q$, radii of the horizon and static limit, as well as the volume of the ergoregion decrease.

\section{Separability of variables in the Hamilton-Jacobi equation for neutral test particle and Carter equations}\label{sec-hj}

The test particles (photons) move along the spacetime geodesics. The motion is governed by the Hamilton-Jacobi equation that reads
\begin{eqnarray}\label{hj0}
g^{\mu\nu}\frac{\partial S}{\partial x^\mu}\frac{\partial S}{\partial x^\nu}=-m^2,
\end{eqnarray}
where $m$ is the mass of the test particle. Due to the stationarity and axial symmetry of the spacetime, we can introduce two integrals of the motion, energy $E = -p_t$, and axial angular momentum $L = p_{\phi}$.
Then we can write the Hamilton-Jacobi action function $S$ in the following separated form
\begin{eqnarray}\label{hj}
S=-\frac{1}{2}m^2\tau-Et+L\phi+S_r(r)+S_\theta(\theta)\ ,
\end{eqnarray}
where $m^2=0, +1$ is considered for null and timelike geodesics, respectively, $\tau$ is the proper time of the test particle with $m^2=+1$. Contravariant components of the metric tensor of the spacetime (\ref{rotating2}) have the form
\begin{eqnarray}\label{contravariant}
&&g^{tt}=-\frac{(r^2+a^2)^2-a^2\Delta\sin^2\theta}{\Delta\Sigma}, \quad
g^{t\phi}=-\frac{2a\rho r}{\Delta\Sigma},\nonumber\\
&&g^{rr}=\frac{\Delta}{\Sigma}, \quad g^{\theta\theta}=\frac{1}{\Sigma}, \quad
g^{\phi\phi}=\frac{\Delta-a^2\sin^2\theta}{\Delta\Sigma\sin^2\theta} .
\end{eqnarray}
Substituting (\ref{hj}) and (\ref{contravariant}) to (\ref{hj0}), we obtain
\begin{eqnarray}\label{05}
&&-\left[\frac{(r^2+a^2)^2}{\Delta}-a^2\sin^2\theta\right]E^2+\frac{4a\rho r}{\Delta}EL\nonumber\\ &&+\left[\frac{1}{\sin^2\theta}-\frac{a^2}{\Delta}\right]L^2+\Delta \left(\frac{\partial S}{\partial r}\right)^2+\left(\frac{\partial S}{\partial \theta}\right)^2+m^2\Sigma=0.\nonumber\\
\end{eqnarray}
Simplifying the equation (\ref{05}), introducing separation constant $K$ representing an additional constant of the motion, and a new constant of the motion through the relation ${\cal Q} = K - (L - aE)$, we arrive to
\begin{eqnarray}\label{06}
&&\Delta\left(\frac{dS}{dr}\right)^2=\frac{R(r)}{\Delta},\\
&&\left(\frac{dS}{d\theta}\right)^2=\Theta(\theta)\ ,
\end{eqnarray}
where
\begin{eqnarray}
&&R(r)=\left[(r^2+a^2)E-aL\right]^2-
\Delta\left[(aE-L)^2+m^2r^2+{\cal Q}\right],\label{06a}\nonumber\\
\\
&&\Theta(\theta)={\cal Q}-\left[\frac{L^2}{\sin^2\theta}+ a^2\left(m^2-E^2\right)\right]\cos^2\theta\ .\label{06b}
\end{eqnarray}
We can write the Hamilton-Jacobi action (\ref{hj}) in terms of these functions as
\begin{eqnarray}\label{hj2}
S=\frac{1}{2}m^2\tau-Et+L\phi+\int^r\frac{\sqrt{R(r)}}{\Delta}+\int^\theta\sqrt{\Theta(\theta)}.
\end{eqnarray}
and obtain the modified Carter equations for the generic regular rotating spacetimes
\begin{eqnarray}\label{eqs}
&&\Sigma\dot{t}=\frac{r^2+a^2}{\Delta}\left[E(r^2+a^2)-aL\right]-a(aE\sin^2\theta-L),\\
&&\Sigma\dot{r}=\sqrt{R}\ ,\\
&&\Sigma\dot{\theta}=\sqrt{\Theta}\ ,\\
&&\Sigma\dot{\phi}=\frac{a}{\Delta}\left[E(r^2+a^2)-aL\right]-\left(aE-\frac{L}{\sin^2\theta}\right)\ ,
\end{eqnarray}
where the overdot ($\dot{ }$) stands for the derivative with respect to the proper time $\tau$ (affine parameter for photons).

\section{Circular orbits around generic rotating regular black holes}\label{sec-circular}

In this section we aim to study the circular motion of the test particles around generic rotating regular black holes. As is well known in the rotating axially symmetric and stationary spacetimes of the type considered here, the circular geodesic orbits have to be confined to the equatorial plane ($\theta=\pi/2$). Therefore, the velocity of the particle along $\theta$ coordinate is zero ($\dot{\theta}=0$), consequently, the constant of motion related to the $\theta$ coordinate $\dot{\cal{Q}}=0$. The circular orbits of the particles around the black hole are determined by the conditions
\begin{eqnarray}\label{cir-con}
R(r)=0, \qquad \frac{dR(r)}{dr}=0.
\end{eqnarray}
By solving the above equations one can find the specific energy and the specific angular momentum of the test particle moving along the circular orbits around the generic rotating regular black hole in the general form
\begin{eqnarray}
&&E=\pm\frac{g_{tt}+g_{t\phi}\Omega}{\sqrt{-(g_{tt}+2g_{t\phi}\Omega+g_{\phi\phi}\Omega^2)}},\label{ener}\\
&&L=\mp\frac{g_{t\phi}+g_{\phi\phi}\Omega}{\sqrt{-(g_{tt}+2g_{t\phi}\Omega+g_{\phi\phi}\Omega^2)}},\label{angul}\\
&&\Omega=\frac{-g_{t\phi,r}\pm\sqrt{g_{t\phi,r}^2-g_{tt,r}g_{\phi\phi,r}}}{g_{\phi\phi,r}},\label{ang-vel}
\end{eqnarray}
where $+$ and $-$ signs represent corotating and counterrotating particles relative to the spacetime rotation, respectively, and $\Omega=d\phi/dt$ is the angular velocity of the particle relative to the distant observers. By using the generic rotating regular black hole ($M=0$) line element~(\ref{rotating2}), we rewrite Eqs.~(\ref{ener})--(\ref{ang-vel}) as follows:
\begin{eqnarray}
&&E=\pm\frac{r-2\rho+a\sqrt{\frac{\rho}{r}-\rho'}}{\sqrt{r\left(r-3\rho+ 2a\sqrt{\frac{\rho}{r}-\rho'}+r\rho'\right)}},\label{ener1}\\
&&L=\mp\frac{2a\rho-(r^2+a^2)\sqrt{\frac{\rho}{r}-\rho'}}{\sqrt{r\left(r-3\rho+ 2a\sqrt{\frac{\rho}{r}-\rho'}+r\rho'\right)}},\label{angul1}\\
&&\Omega=\frac{\sqrt{\frac{\rho}{r}-\rho'}}{a\sqrt{\frac{\rho}{r}-\rho'}\pm r}.\label{ang-vel1}
\end{eqnarray}
From Eqs.~(\ref{ener1})--(\ref{ang-vel1}) one can see that for the particles moving along the circular orbits the following conditions must be satisfied:
\begin{eqnarray}
&&\frac{\rho}{r}-\rho'\geq0\ ,\label{st-rad1}\\
&&r-3\rho+ 2a\sqrt{\frac{\rho}{r}-\rho'}+r\rho'>0\ .\label{st-rad10}
\end{eqnarray}
By simplifying (\ref{st-rad10}), one can see that the solution of inequality~(\ref{st-rad10}) fully satisfies the condition~(\ref{st-rad1}). Therefore, we conclude that the condition for existence of circular orbits around the generic rotating regular black hole~(\ref{rotating2}) (for $r>0$) reads
\begin{eqnarray}\label{st-rad2}
\rho-r\rho'\geq0\ ,
\end{eqnarray}
or $f'(r)>0$. For the generic rotating regular black holes one can write the condition~(\ref{st-rad2}) in the more explicit form
\begin{eqnarray}\label{st-rad-gen}
r\geq q\left(\mu-1\right)^{1/\nu}\ .
\end{eqnarray}
Interestingly, the rotation of the black hole does not play a role in the condition giving limit on the existence of the circular orbits. One can see from~(\ref{st-rad2}) that in the Kerr spacetime ($\rho=M$) $M/r>0$ is the limiting condition for existence of circular orbits; the circular orbits can approach both zero and infinity; of course there exists also limits given by circular photon orbits. In the case of the Kerr-de Sitter~\cite{StuchlikPRD:2004} or quintessential~\cite{Tos:Arxiv:2015} black holes, the spacetimes have de-Sitter behaviour at infinity (at large distances) and consequently, there exists the static radius limit at large distances depending on the value of the field parameter and circular orbits can exist below the static radius ($r<r_s$)~\cite{StuchlikBAC:1983}. In the case of the rotating regular black holes, on no-horizon spacetimes the spacetimes have de-Sitter-like behavior near the center. Therefore, there exists a static radius limit near the center, and the circular orbits can appear at larger distances from it ($r>r_s$).

From the astrophysical point of view, the most relevant among the circular orbits are those giving limits governing the optical appearance and accretion phenomena, i.e., the photon circular orbits and marginally stable circular geodesics.  In the photon circular orbit the specific energy and angular momentum of the particles diverge. So from (\ref{ener1}) and (\ref{angul1}) we can get the expression for the photon orbit as
\begin{eqnarray}\label{photon1}
r\left[1-(r^2+a^2)\Omega^2\right]-2(1-a\Omega)^2\rho=0\ .
\end{eqnarray}
By inserting~(\ref{ang-vel1}) to the  above equation and solving it with respect to the rotation parameter, we obtained the condition of the loci of the photon orbits around generic rotating regular black holes in the form
\begin{eqnarray}\label{photon2}
a_{ps}^2\equiv\frac{r(r-3\rho+r\rho')^2}{4(\rho-r\rho')}.
\end{eqnarray}

Stable circular orbits are defined by the condition
\begin{eqnarray}\label{mar-stab}
\frac{d^2R(r)}{dr^2}\geq0
\end{eqnarray}
that has to be satisfied simultaneously with the conditions~(\ref{cir-con}). Marginally stable circular orbits are determined by the equality in the above given condition. Due to the cumbersome form of the expression of the marginally stable orbits or innermost stable circular orbit (ISCO), we write it as follows:
\begin{eqnarray}\label{isco}
a_{ms\pm}^2\equiv\frac{A\pm16 \sqrt{B}}{C}\ ,
\end{eqnarray}
where
\begin{eqnarray}\label{coefs}
A&&=2r\left[14\rho^3+\rho^2(3r-48r\rho')+2r^2\rho(27\rho'^2\right.\nonumber\\ &&\left.+r\rho''(r\rho''-1)+\rho'(2r\rho''-3))-r^3(20\rho'^3\right.\nonumber\\ &&\left.-2r\rho'\rho''+r^2\rho''^2+\rho'^2(4r\rho''-3))\right],\nonumber\\
B&&=r^2(r\rho'-\rho)^3\left[-3r\rho^2+2\rho^3-2r^2\rho(3\rho'^2\right.\\ &&\left.+r\rho''(r\rho''-1)+\rho'(2r\rho''-3))+r^3(4\rho'^3\right.\nonumber\\ &&\left.-2r\rho'\rho''+r^2\rho''^2+\rho'^2(4r\rho''-3))\right],\nonumber\\
C&&=2\left[3\rho+r(r\rho''-3\rho')\right]^2.\nonumber
\end{eqnarray}

From the expression (\ref{isco}) one can see that to have stable circular orbits the condition $B\geq0$ must be satisfied together with the condition for existence of circular orbits~(\ref{st-rad2}). Then, we arrive to the condition
\begin{eqnarray}\label{con-1}
&&-3r\rho^2+2\rho^3-2r^2\rho(3\rho'^2+r\rho''(r\rho''-1)+\rho'(2r\rho''-3))\nonumber\\ &&+r^3(4\rho'^3-2r\rho'\rho''+r^2\rho''^2+\rho'^2(4r\rho''-3))\leq0\ ,
\end{eqnarray}
We concentrate our attention to the special case of regular spacetimes representing the physically interesting limit.

\subsection{Circular orbits around rotating Maxwellian regular black hole}

In this subsection we study in detail the circular orbits of the test particles around the rotating regular black holes related to the Maxwell weak field limit of the nonlinear electrodynamics, i.e., the new type of regular black hole classes with $\nu=1$~(\ref{new1})\footnote{In this case $\rho(r)=M+\frac{r^\mu}{(r+q)^\mu}$. Since we are considering the regular black hole, $M=0$.}. In this case the condition for existence of the circular orbits~(\ref{st-rad-gen}) is written in the form
\begin{eqnarray}\label{st-rad3}
r_s\geq q(\mu-1)=\frac{2Q_{em}^2}{M_{em}^2}(\mu-1)\ .
\end{eqnarray}
If we take the minimum value $\mu=3$ for the black hole to be regular and compare the result, $r\geq2q$, with Fig.~\ref{fig-hor}, we obtain the amazing result that this condition is the same as the existence condition of the black holes~(\ref{extreme3}) in the case of $a=0$. This condition is satisfied in the region of the generic rotating black hole too (see the last plot of the top panel of Fig.~\ref{fig-hor}).

For the case of $\nu=1$, the location of the circular photon orbit can be expressed by the relation
\begin{eqnarray}\label{photon3}
&&a^2=a_{ps}^2\equiv\frac{\left[r(r+q)^{\mu+1}+r^{\mu}(q\mu-3(r+q))\right]^2} {4r^{\mu-1}(r+q-q\mu)(r+q)^{\mu+1}},
\end{eqnarray}
In the special case of $\mu=3$, we obtain the condition for the radius of the photon circular orbit in the form
\begin{eqnarray}\label{photon4}
a_{ps}^2\equiv\frac{\left[q^2(6r^2+4qr+q^2)+(4q-3)r^3+r^4\right]^2}{4(r-2q)(r+q)^4}.
\end{eqnarray}
In order to study the stability of the circular orbits around the rotating Maxwellian regular black hole we write the inequality~(\ref{con-1}) in the form
\begin{figure*}[ht]
\begin{center}
\includegraphics[width=0.32\textwidth]{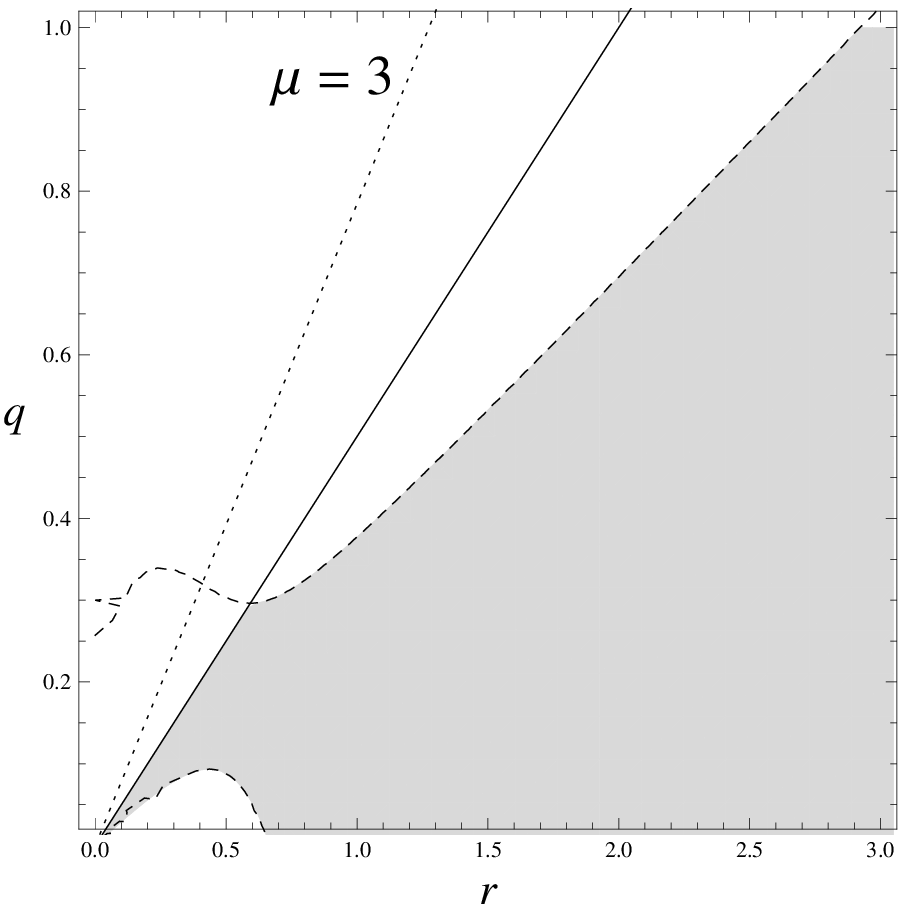}
\includegraphics[width=0.32\textwidth]{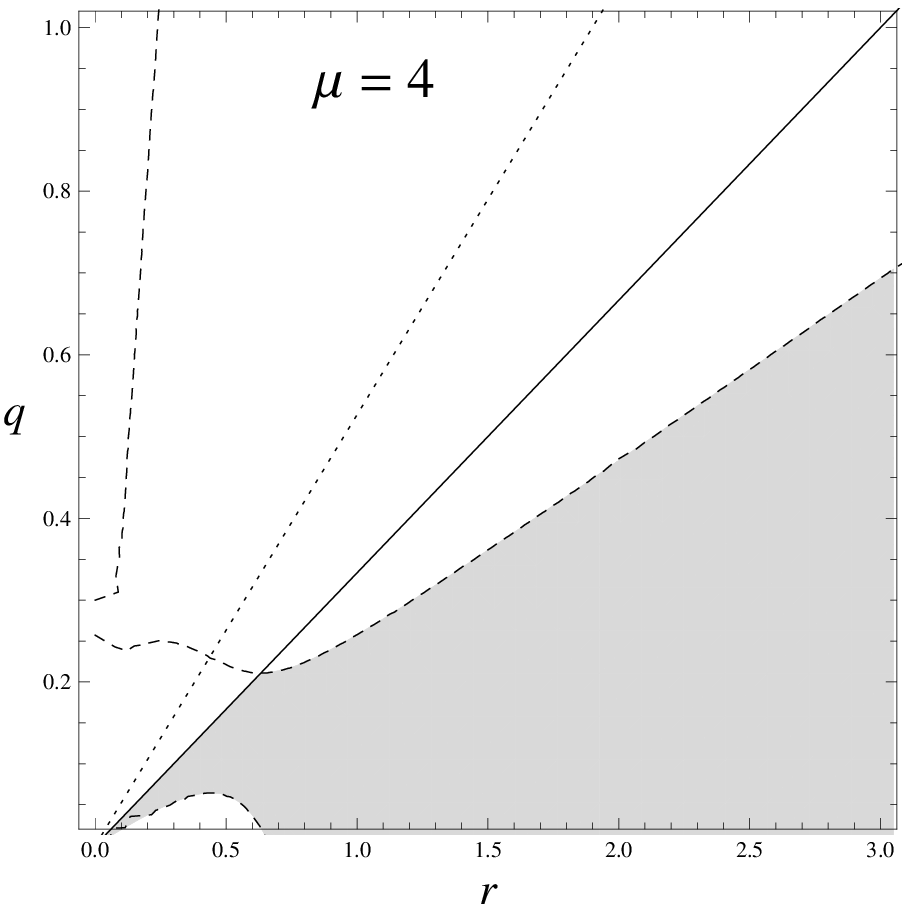}
\includegraphics[width=0.32\textwidth]{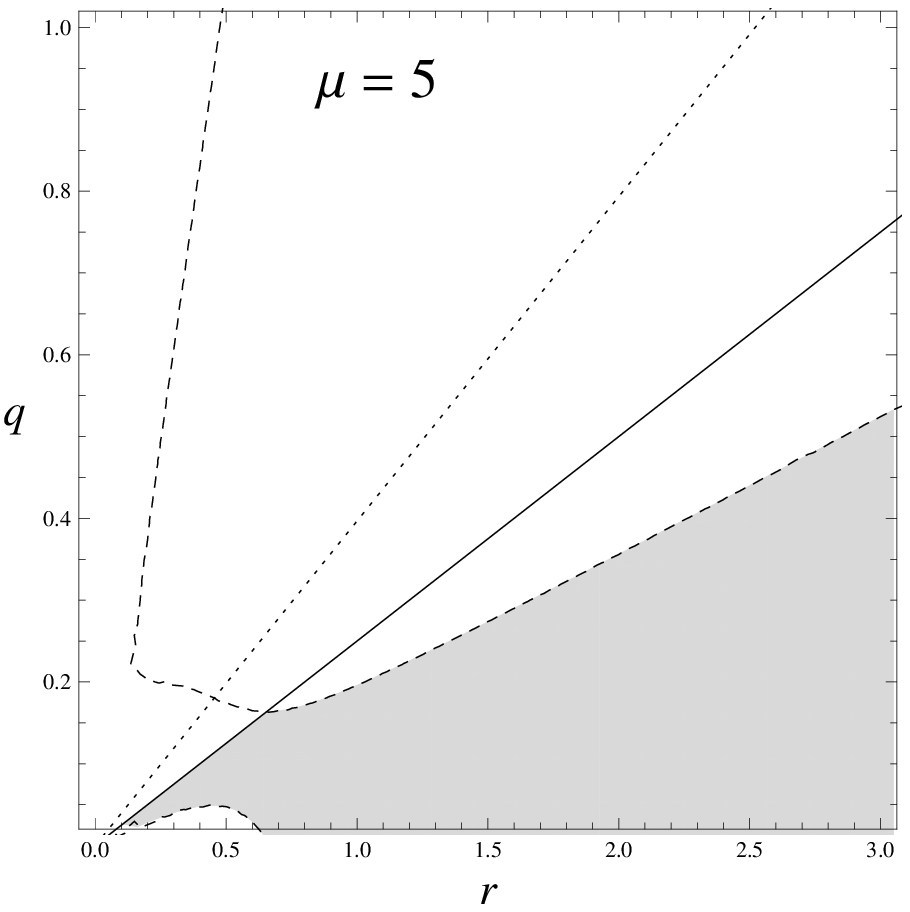}
\end{center}
\caption{\label{fig-stab} Condition for the existence of the stable circular orbits in parametric space of the generic rotating new type of regular black hole with $\nu=1$ for different values of $\mu$. Where solid, dashed and dotted lines represents the static radius $q_{s}$, lower $q_{ms-}$ and upper $q_{ms+}$ limit of the condition for marginally stable circular orbits, respectively. The stable circular orbits can exist in shaded region.}
\end{figure*}
\begin{eqnarray}\label{con-2}
F(r,q_{ms-}(r))(q-q_{ms+}(r))\leq0\ ,
\end{eqnarray}
with
\begin{eqnarray}\label{qm+}
q_{ms+}(r)=\frac{\mu+2+\sqrt{\mu(5\mu+4)}}{2(\mu^2-1)}r\ ,
\end{eqnarray}
and
\begin{eqnarray}\label{Fqm}
F&&=2r^\mu((r+q)^2-q\mu(q\mu+r))\\
&&-r(r+q)^\mu(3r^2+qr(6-5\mu)+q^2(\mu-3)(\mu-1))\ ,\nonumber
\end{eqnarray}
The condition~(\ref{st-rad2}) ($q-r/2<0$) guarantees the expression $q-q_{ms+}(r)$ to be negative ($q-q_{ms+}(r)<0$). Therefore, the condition $F(q,q_{ms-}(r))>0$ must be satisfied. In Fig.~\ref{fig-stab} we summarize all conditions and give the parametric region for existence of stable circular orbit for the different values of parameter $\mu$. One can see from Fig.~\ref{fig-stab} that stable circular orbits are restricted only by the conditions $q_{s}$ and $q_{ms-}$. In the case of $\mu=3$ these curves intersect at $q\approx0.296$ and $r\approx0.593$ which represents the non-rotating extreme new type of regular black hole (see Fig.~\ref{fig-hor}). Starting from this point ($q\geq0.593$), the spacetimes represent the rotating no-horizon spacetimes. Comparing Figs.~\ref{fig-hor} and~\ref{fig-stab}, one can deduce that for the no-horizon spacetimes the stable circular orbits can exist everywhere, even near the center. With increasing the value of $\mu$, the parametric region for the existence of stable circular orbits decreases. A detailed study of the no-horizon rotating regular spacetimes is postponed for a future paper.
\begin{figure*}[t]
\begin{center}
\includegraphics[width=0.32\linewidth]{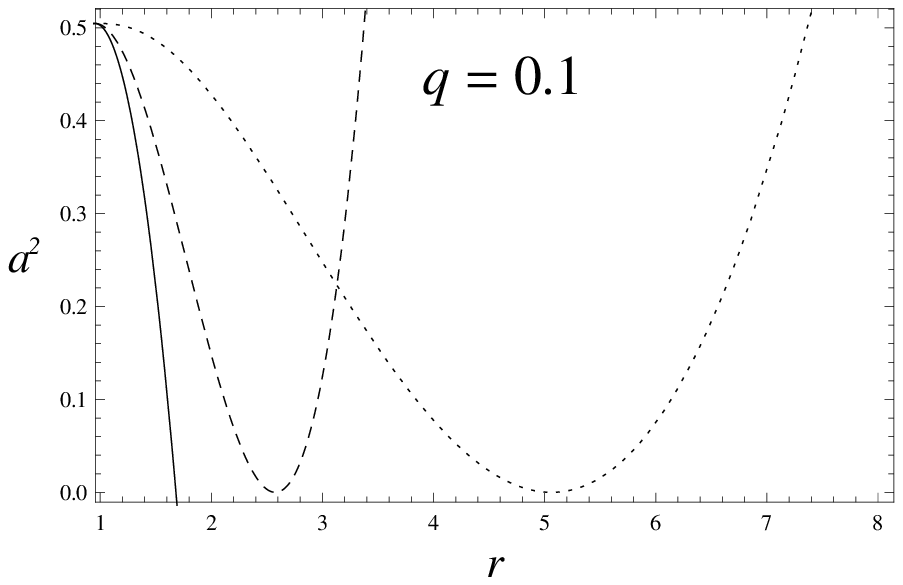}
\includegraphics[width=0.32\linewidth]{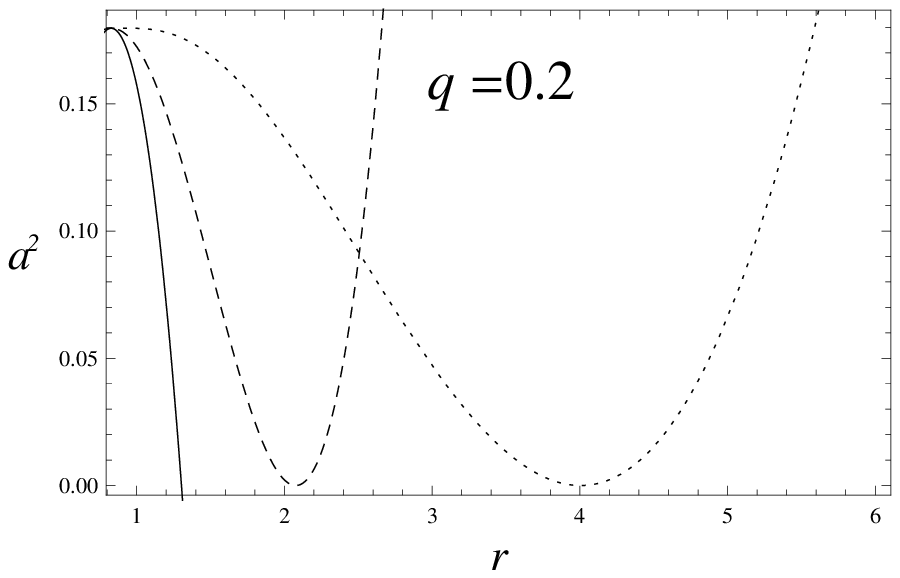}
\includegraphics[width=0.32\linewidth]{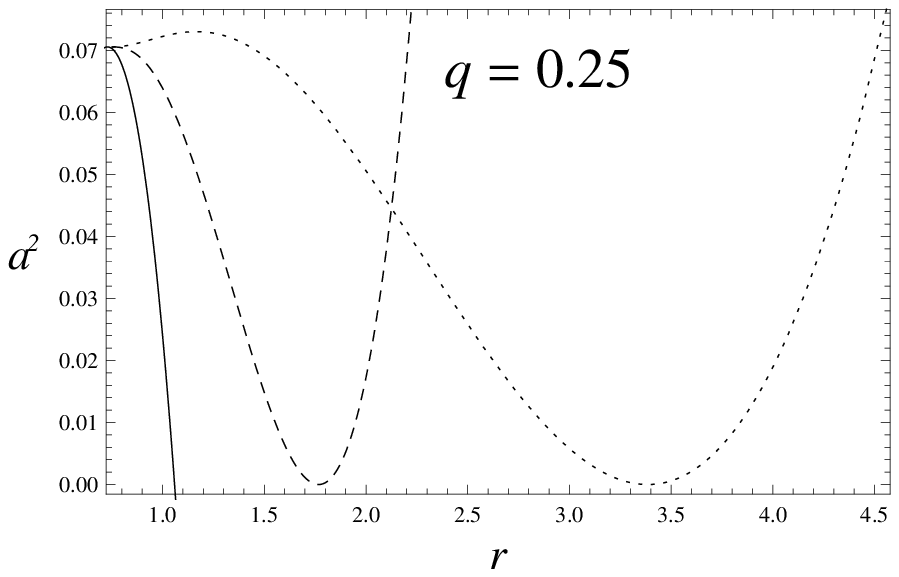}
\end{center}
\caption{\label{fig-orbit} Locations of horizon (solid), photon orbit (dashed) and ISCO (dotted) for different values of the magnetic (electric) charge parameter with $\nu=1$, $\mu=3$.}
\end{figure*}

In Fig.~\ref{fig-orbit} we show illustrative plots for the location of the main characteristic radii of the circular orbits: horizon, photon orbit and marginally stable orbits are given for different values of the charge parameter $q$ of the new type of the rotating regular black hole with $\nu=1$, $\mu=3$. One can see from Fig.~\ref{fig-orbit} that with increasing value of the specific charge parameter not only radii of all characteristic orbits, but also the possible values of the rotation parameter of the black hole for the existence of these circular orbits decreases as well.

\section{Harmonic oscillations of neutral test particle}\label{sec-epicyclic}

It is known that the stable circular orbits of the neutral test particles are located at the minima of the radial function $R$ (\ref{06a}). If particles deviate slightly from the stable circular orbit it starts to oscillate around its equilibrium realizing an epicyclic motion. These oscillating motions of the particle are governed by the epicyclic frequencies. Let us consider the deviations of the particle from the stable circular orbit at $r_0$ and $\theta_0=\pi/2$ which corresponds to the minimum of the radial function $R$. The radii of the oscillatory motion are denoted as $r=r_0+\delta r$ and $\theta=\theta_0+\delta\theta$. Then, evolution of these deviations is described by the equations~\cite{AbramowiczAPSS:2005,TursunovPRD:2016}
\begin{eqnarray}\label{epic}
\ddot{\delta r}+\omega_r^2\delta r=0, \qquad \ddot{\delta \theta}+\omega_\theta^2\delta \theta=0,
\end{eqnarray}
where the overdot stands for the derivative with respect to the proper time of the particle, $\omega_r$ and $\omega_\theta$ are locally measured radial and latitudinal epicyclic frequencies with respect to the proper time of
a comoving observer. They are defined by the relations~\cite{AbramowiczAPSS:2005}
\begin{eqnarray}\label{epic-th}
\omega_r^2=-\frac{1}{2g_{rr}}\frac{\partial^2V_{eff}}{\partial r^2},\quad
\omega_\theta^2=-\frac{1}{2g_{\theta\theta}}\frac{\partial^2V_{eff}}{\partial \theta^2},
\end{eqnarray}
where the effective potential $V_{eff}$ is given by the particle circular motion parameters $E$, $L$ through the relation
\begin{eqnarray}\label{veff}
V_{eff}=\frac{g_{\phi\phi}E^2+2g_{t\phi}EL+g_{tt}L^2}{g_{t\phi}^2-g_{tt}g_{\phi\phi}}-1.
\end{eqnarray}
Moreover, there is another very important angular frequency of the circular epicyclic motion of the particle, namely the azimuthal frequency of the circular motion in the equatorial plane, defined by the relation
\begin{eqnarray}
\omega_\phi=\dot{\phi}=\frac{2aE\rho+L(r-2\rho)}{r\Delta}.
\end{eqnarray}

Of course, astrophysically relevant are the epicyclic frequencies related to the distant static observers. These are given by the frequencies related to the proper time of the oscillatory particle, modified by the redshift factor
\begin{eqnarray}\label{redshift}
\dot{t}=\frac{r+a\sqrt{\frac{\rho}{r}-\rho'}}{\sqrt{r\left(r-3\rho+ 2a\sqrt{\frac{\rho}{r}-\rho'}+r\rho'\right)}}.
\end{eqnarray}
The radial and vertical epicyclic frequencies with respect to the coordinate time $t$ related to the observers at rest very far from the source are then given by the relations
\begin{eqnarray}\label{epic-distant}
\Omega_r^2=\frac{\omega_r^2}{\dot{t}^{2}}, \qquad \Omega_\theta^2=\frac{\omega_\theta^2}{\dot{t}^{2}}.
\end{eqnarray}
The Keplerian (azimuthal) frequency $\Omega_\phi$ is given by the relation~(\ref{ang-vel1}) and for the corotating particle it reads
\begin{eqnarray}\label{keplerian}
\Omega_\phi=\frac{\sqrt{\frac{\rho}{r}-\rho'}}{r+a\sqrt{\frac{\rho}{r}-\rho'}}.
\end{eqnarray}

\begin{figure*}[t]
\begin{center}
\includegraphics[width=0.32\linewidth]{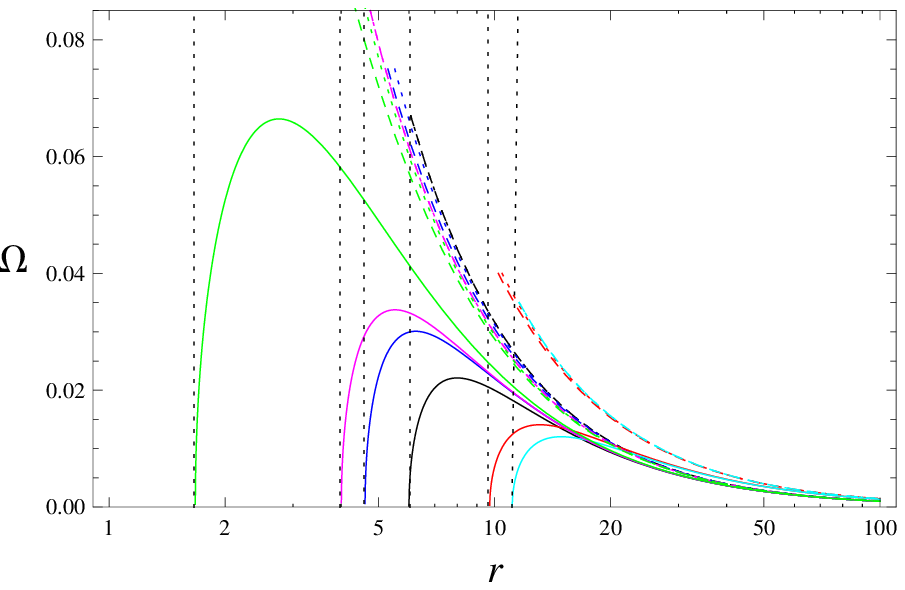}
\includegraphics[width=0.32\linewidth]{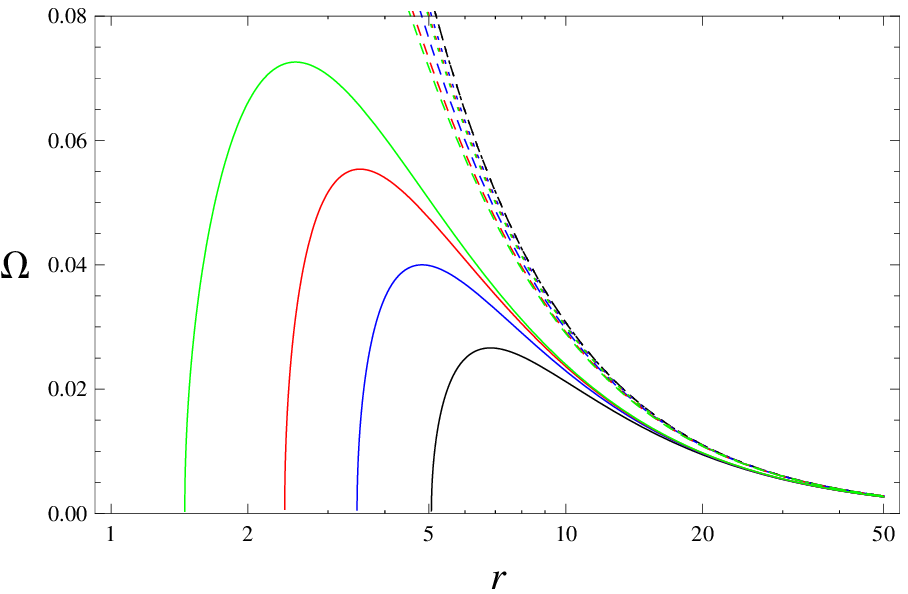}
\includegraphics[width=0.32\linewidth]{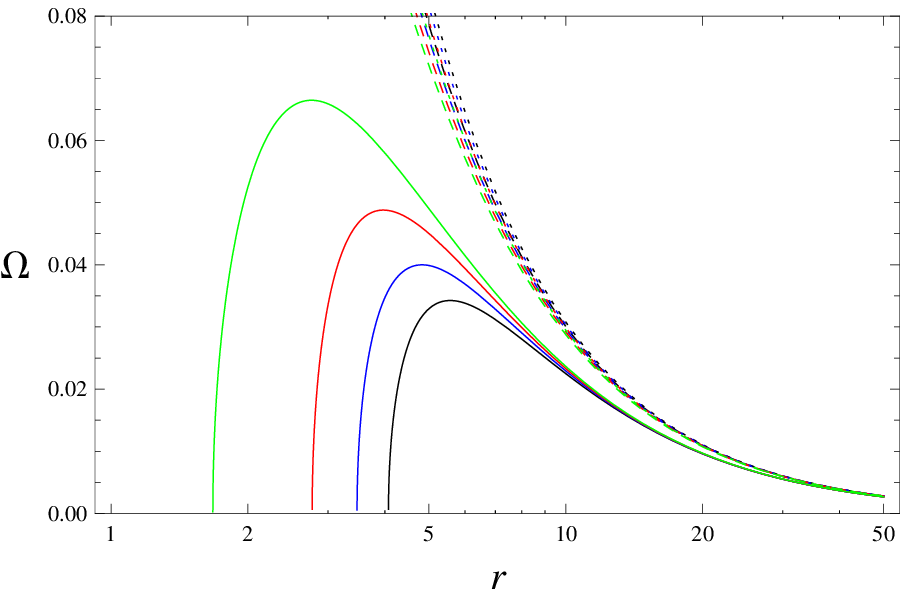}
\end{center}
\caption{\label{fig-epicyclic} Radial profiles of epicyclic frequencies measured by the distant observer ($\Omega_r$ -- solid, $\Omega_\theta$ -- dashed, $\Omega_\phi$ -- dotted curves) for the neutral test particle moving along the stable circular orbit around: \textit{\textbf{Left panel:}} The Schwarzschild black hole -- black, the Kerr black hole with $a/M=0.4$ -- blue, non-rotating Maxwellian singular black hole with $\mu=3$ and $q/M=0.2$ -- cyan, rotating Maxwellian singular black hole with $\mu=3$, $a/M=0.4$ and $q/M=0.2$ -- red, non-rotating Maxwellian regular black hole with $\mu=3$ and $q=0.2$ -- magenta, rotating Maxwellian regular black hole with $\mu=3$, $a=0.4$, and $q=0.2$ -- green. Vertical dotted lines represent the loci of ISCO. \textit{\textbf{Middle panel:}} Rotating regular Maxwellian black hole with $\mu=3$ and $q=0.1$ for different values of the specific charge parameter: $a=0$ -- black, $a=0.4$ -- blue, $a=0.6$ -- red, $a=0.7$ -- green. \textit{\textbf{Right panel:}} The same as the middle panel but for the fixed rotation parameter $a=0.4$ and different values of the specific charge parameter: $q=0.05$ -- black, $q=0.1$ -- blue, $q=0.15$ -- red, $q=0.2$ -- green.}
\end{figure*}

By inserting~(\ref{veff}),~(\ref{epic-th}) and~(\ref{redshift}) into Eq.~(\ref{epic-distant}) we obtain the radial and latitudinal epicyclic frequencies of the neutral particle in the epicyclic motion around the circular equatorial orbits ($\theta=\pi/2$) in the form
\begin{widetext}
\begin{eqnarray}
\Omega_r^2&&=\frac{[r(L^2-E^2(r^2+a^2))-2(aE-L)^2\rho]\Delta'^2}{r[Er(r^2+a^2)+2a(aE-L)\rho]^2}- \frac{\Delta^2[E^2r^3+(aE-L)^2(2\rho+r(r\rho''-2\rho'))]}{r^3[Er(r^2+a^2)+2a(aE-L)\rho]^2}\nonumber\\
&&-\frac{2\Delta[-E^2r^3+(aE-L)^2(\rho-r\rho')]\Delta'}{r^2[Er(r^2+a^2)+2a(aE-L)\rho]^2} +\frac{\Delta[r(-L^2+E^2(r^2+a^2))+2(aE-L)^2\rho]\Delta''}{2r[Er(r^2+a^2)+2a(aE-L)\rho]^2},\\
\Omega_\theta^2&&=\frac{\Delta[L^2r^3+2\rho(aE-L)(a^2(aE-L)+r^2(aE+L))]}{r^3[Er(r^2+a^2)+2a(aE-L)\rho]^2},
\end{eqnarray}
\end{widetext}
where $E$ and $L$ represent the values of energy and angular momentum of the particle moving along the stable circular orbits, given by Eqs.~(\ref{ener1})--(\ref{ang-vel1}). It is worth noting that the radial epicyclic frequency vanishes at ISCO. We derive the epicyclic frequencies measured by the distant observer for the particles moving along the circular stable orbits around several black holes:
\begin{itemize}

\item $\rho=M$, $a=0$ -- Schwarzschild spacetimes;

\item $\rho=M$ -- Kerr spacetimes;

\item $\rho=M+\frac{r^\mu}{(r+q)^\mu}$, $a=0$ -- non-rotating Maxwellian singular spacetimes;

\item $\rho=\frac{r^\mu}{(r+q)^\mu}$, $a=0$ -- non-rotating Maxwellian regular spacetimes;

\item $\rho=M+\frac{r^\mu}{(r+q)^\mu}$ -- rotating Maxwellian singular spacetimes;

\item $\rho=\frac{r^\mu}{(r+q)^\mu}$ -- rotating Maxwellian regular spacetimes.

\end{itemize}
%
%
%\begin{eqnarray}
%\omega_r^2&&=\frac{3(\Delta-a^2)(L-aE)^2-E^2r^4}{r^6}\nonumber\\ &&-\frac{\left[(r^2+a^2)E-aL\right]^2\Delta'^2}{r^4\Delta^2}\nonumber\\
%&&-\frac{2\left[a^2(L-aE)^2-E^2r^4\right]\Delta'}{r^5\Delta^2}\nonumber\\ &&+\frac{\left[(r^2+a^2)E-aL\right]^2\Delta''}{2r^4\Delta^2}\ ,\label{epic-r2}\\
%\omega_\theta^2&&=\frac{a^2\left[(r^2+a^2)E-aL\right]^2}{r^6\Delta^2}\nonumber\\
%&&-\frac{a^2(L-aE)^2+r^2(a^2E^2-L^2)}{r^6\Delta}\ ,\label{epic-th2}
%\end{eqnarray}
%
In order to compare the epicyclic frequencies measured by distant observer for the above--mentioned spacetimes, we show the behaviors of them in the left panel of Fig.~\ref{fig-epicyclic}. In Fig.~\ref{fig-epicyclic} we present the behavior of the epicyclic frequencies measured by distant observers for the oscillatory motion of the neutral test particle moving along the epicyclic orbits around rotating Maxwellian regular black holes. We give the radial profiles of the epicyclic frequencies for characteristic values of charge parameter $q$ and spin $a$.
Near the ISCO, the effect of the gravitational field of the black hole is strong and the $\Omega_\phi$ and latitudinal $\Omega_\theta$ epicyclic frequencies are almost equal for the small values of rotation parameter $a$ and they are always much bigger than the radial $\Omega_r$ epicyclic frequency, $\Omega_\theta, \Omega_\phi\gg\Omega_r$. At large distances all characteristic frequencies tend to zero due to weakening of the gravitational field.

From the left panel of Fig.~\ref{fig-epicyclic} one can deduce that rotation of the black hole pulls the ISCO toward the black hole and increases the epicyclic frequencies. However, the charge parameters of the Maxwellian regular and singular black hole spacetimes play an inverse role in comparison to the role of the rotation parameter, i.e., it increases the ISCO and decreases the values of the epicyclic frequencies (see the right panel of Fig.~\ref{fig-epicyclic}). For example, in both the rotating Maxwellian singular ($\Omega_{r,Ms}$) and regular ($\Omega_{r,Mr}$) black hole spacetimes, the values of the radial epicyclic frequencies measured by the distant observer cannot be as big as in the Kerr black hole spacetime ($\Omega_{r,K}$), while the latitudinal and azimuthal epicyclic frequencies do not differ substantially. When the Maxwellian singular and regular black holes have the same values of the charge $q$ and rotation $a$ parameters, the epicyclic frequencies in the regular one with smaller ISCO radius is bigger than in the singular one -- see the left panel of Fig.~\ref{fig-epicyclic}. Thus, the comparisons of the epicyclic frequencies in the above--mentioned spacetimes have shown that $\Omega_{r,K}>\Omega_{r,Mr}>\Omega_{r,Ms}$.

\section{Conclusion}\label{sec-conc}

In the present paper we introduce the generic rotating black hole solutions in general relativity coupled to the nonlinear electrodynamics. We thus generalize the spherically symmetric black hole solution obtained by Fan and Wang~\cite{Fan-Wan:2016:arxiv1610.02636:} using the method based partly on the Newman-Janis algorithm. In our model one can construct exact black hole solutions by choosing appropriate values for the characteristic parameters $\mu$ and $\nu$ as in the case of the nonrotating case, when the known solutions can be obtained for fixed values of the parameters: $\nu=2$ -- Bardeen-like black holes, $\nu=3$ -- Hayward-like black holes, and $\nu=1$ -- new type of black hole solution which approaches the Maxwell field in a weak field limit. Moreover, we have shown that some main properties of these rotating black holes are almost the same as those occurring in the nonrotating case which was discussed in~\cite{Fan-Wan:2016:arxiv1610.02636:}. Especially, in the case of zero gravitational mass black hole, $M=0$, the $\mu\geq3$ spacetimes are the regular (nonsingular) ones. Furthermore, we have shown that the presence of the gravitational mass does not affect the energy conditions and the fact that the obtained solutions violate the WEC and SEC. Though an increase in the value of $\mu$ decreases the depth of the violation of energy conditions, the violations are always preserved even for the large values of $\mu$.

It has been shown that as the standard (rotating) Bardeen, Hayward and ABG regular black holes there is a upper limit on the value of the charge parameter for existence of horizon, i.e., spacetime to represent the black hole. For greater values of the charge parameter than these critical values, $q>q_{cr}$, spacetimes represent no-horizon ones. The rotating singular black hole with gravitational mass $M$ has always two horizons for $a/M<1$ irrespective of the value of the charge parameter $q$ and with increasing the value of the charge parameter $q$ the radius of horizon increases.

Furthermore, we have shown that these new solutions can be written in the Kerr-like form and give separable Hamilton-Jacobi equations of the geodesic equations for motion of neutral test particles. Circular orbits of the test particle around rotating regular black holes have been studied and it has been shown that the condition for existence of the circular orbits (static radius $r_s$) does not depend on the rotation of the black hole as in the Kerr, Kerr-Newman, Kerr-de Sitter and quintessential Kerr black holes. However, unlike the case of the Kerr-de Sitter~\cite{StuchlikPRD:2004} and quintessential Kerr~\cite{Tos:Arxiv:2015} black holes, where $r_{circ}<r_s$, in the field of these rotating regular black holes the static radius is located near the black hole and circular orbits can exist above the static radius, $r>r_s$. As a special case we have shown the characteristic circular orbits, namely, the photon orbit and ISCO around the new type of regular black holes where the nonlinear electrodynamic field tends to the Maxwell field in the weak field limit. We have discussed the stable circular orbits in the field of the Maxwellian rotating regular black hole and give the radial and vertical frequencies of the epicyclic motion. The character of the radial profiles of the epicyclic frequencies indicates possibility to find signatures of the Maxwellian rotating regular black holes in high frequency quasiperiodic oscillations observed in some microquasars.

\begin{acknowledgments}

The authors would like to thank the anonymous referees for their careful reading of our manuscript and many insightful comments and suggestions that have improved the paper. B.T. and Z.S. would like to express their acknowledgments for the institutional support of the Faculty of Philosophy and Science of the Silesian University in Opava, the internal student grant of the Silesian University (Grant No. SGS/14/2016) and the Albert Einstein Centre for Gravitation and Astrophysics under the Czech Science Foundation (Grant No.~14-37086G). B.A. acknowledges the Faculty of Philosophy and Science, Silesian University in Opava, Czech Republic, and the Goethe University, Frankfurt am Main, Germany, for their warm hospitality. The research of B. A. is supported in part by Grant No.~VA-FA-F-2-008 of the Uzbekistan Agency for Science and Technology, and by the Abdus Salam International Centre for Theoretical Physics through Grant No.~OEA-NT-01 and by the Volkswagen Stiftung, Grant No.~86 866.

\end{acknowledgments}

\end{document}